\documentstyle[12pt,epsf]{article}

\def\bea{\begin{eqnarray}}
\def\eea{\end{eqnarray}}
\def\dst{\displaystyle\phantom{|}}
\def\be{\begin{equation}}
\def\ee{\end{equation}}
\def\ave#1{\langle #1 \rangle}

\def\bk{\mathbf{K}}
\def\bdk{\mathbf{\Delta k}}
\def\bbet{{\mathbf{\beta}}}

\def\D{\Delta}

\def\ov{\over\displaystyle\phantom{|}}
\def\dst{\displaystyle\phantom{|}}
\def\l({\left(}
\def\r){\right)}
\def\ts{\tilde s_c}

\begin{document}
\title{ {\Large\bf Testing the Core-Halo Model \\
	on Bose-Einstein Correlation Functions}}
\author{ S. Nickerson$^1$, T. Cs\"org\H o$^{2}$ and D. Kiang$^1$ \\
	{\normalsize \phantom{HH} } \\
	{\normalsize\it $^1$Department of Physics, 
	Dalhousie University,}\\
	{\normalsize\it  Halifax, N.S., Canada B3H 3J5 }\\
	{\normalsize\it $^2$MTA KFKI RMKI, 
	H-1525 Budapest 114, POB 49, Hungary}}
\date{\null}
\begin{titlepage}
\thispagestyle{empty}
\maketitle
\bigskip
\abstract{
  Having first performed a Monte Carlo simulation to justify the 
analysis technique to be used, we then analyze the Bose-Einstein correlation 
functions from CERN experiment NA44 in the context of the 
core-halo model.  Although experimental resolution {\it and} 
error bar distribution prevents a direct 
observation of the halo structure, the 
values for the core radius and the core fraction of pions can be 
obtained in a straight-forward manner. These are
found to be {\it independent} of the structure of the correlation 
function at small relative momenta of $Q < 50$ MeV. Hence, 
the $\omega$ meson decay products do not distort the Gaussian shape 
of the Bose-Einstein  correlation function in 
$S + Pb \rightarrow \pi + \pi + X$ reaction at CERN SPS. 
As we find that the ``model-independent" HBT radii yield results 
that are quantitatively as well as qualitatively unreliable for systems
with long-lived resonances, we present their corrected form
that applies for correlation functions with $\lambda({\bf K}) < 1$. 
}
\vfill\eject
\end{titlepage}
\section{Introduction}\label{s:ch-int}
High energy heavy ion collisions are providing a reproducible experimental
 environment to study physics at extremely high energy densities 
 in relatively large volumes.
 A tool to access the 
space-time characteristics of such systems is the technique known
as intensity interferometry. If the observed particles are
bosons, intensity interferometry is frequently referred to
as HBT effect~\cite{HBT}  or 
Bose-Einstein correlations~\cite{GGLP0,GGLP,gyu_ka}. 
These intensity correlations
appear due to the enhanced (decreased) likelihood that a boson (fermion) 
is produced in a quantal state that is close in phase-space to
a state already occupied by another boson (fermion).
Intensity interferometry is a fastly developing tool for studying
spatial and temporal extension of hot and dense strongly interacting
systems on the $10^{-15}$ m and $10^{-23}$ s scales.

Accumulating evidence indicates, that high energy heavy ion reactions,
performed currently at CERN SPS and Brookhaven AGS, create composite sources of particles, that can be approximately divided into two parts:
a central core, where the time evolution of the matter can be 
thought as a violent cascading of binary collisions or hydrodynamical
behavior of strongly interacting matter, and a halo of long-lived resonances,
that are created in the hot and dense medium but decay far outside
the core. In case of pions, the resonance halo undoubtedly contains
the decays of $K_S^0$ where $\tau = 8.9 \cdot 10^{-9} $ s,
the decays of the $\eta $ and $\eta^{\prime}$ resonances 
that have a very narrow decay widths of $\Gamma_{\eta} =
1.20 $ keV and $\Gamma_{\eta^{\prime}} = 0.2 MeV$. The lifetimes of these
resonances correspond to $\tau = \hbar / \Gamma = 164 417 $ fm/c and
986.5 fm/c, respectively. The next long-lived resonance which is 
expected to be produced in significant numbers in high energy collisions is
the $\omega $ meson with  a full width of $\Gamma_{\omega} = 8.4 $
MeV and a corresponding  life-time of $\tau_{\omega} = 23.4 $ fm/c.
In the literature, one may observe a scientific debate weather the
decay products of the $\omega$ mesons contribute to the halo or not.

After a brief summary of the statements made about the contribution
of the omega resonance to the BECF-s, we attempt to conclude this debate
in case of the 
$S + Pb \rightarrow \pi + \pi + X$ reaction measured by the 
NA44 Collaboration at CERN SPS~\cite{highpt}, 
by presenting an analysis of this data set along the lines suggested in
ref.~\cite{chalo}. In case of other
reactions or other experimental resolutions/techniques, we emphasize 
that  the role of the
$\omega$ meson should be investigated {\it again}, 
in a manner similar to the analysis presented in the forthcoming sections.  

The main purpose of the present paper is to test, whether the core/halo 
model interpretation of the two-particle correlation function is 
indeed applicable to the above cited NA44 data, or not,
with the help of a novel method suggested in ref.~\cite{chalo}.
If the core/halo interpretation of the measured correlation function
is meaningful, the parameterization of the correlation data should be
independent of the data points at the lowest values of the relative 
momentum, hence these data points could be deleted without significantly
changing the fitted parameters of the correlation functions. 
We check the validity of this hypothesis by gradually removing the
data points at smallest relative momentum from the NA44 $S + Pb$ data set.
To our best knowledge, this is the first time that such kind of 
analysis is performed on measured correlation data, in order to verify the
applicability of the core/halo model.
Hence, we must establish the limits of the  
technique  to be used before attempting the data analysis.

The paper is structured as follows:
In Section~\ref{s:ch-BE} we summarize the basics of Bose-Einstein
correlations and the assumptions that are introduced to obtain the core/halo
model. We   also summarize here those various theoretical considerations
and conclusions about the influence of the $\omega$ meson on the 
two-pion correlation function, that were achieved in earlier numerical
investigations, reviewing the status of this scientific debate. 

In Section~\ref{s:ch-MC}, we describe a Monte-Carlo procedure that was used
to estimate the limits of the procedure that we utilized to 
verify the applicability of the core/halo picture to NA44 data
in the subsequent Section~\ref{s:ch-anal}.
Finally, we summarize and conclude.
Conceptually, our study is also related to a recently discussed
``model-independent" Gaussian parameterization of the correlation 
functions. In Appendix A we evaluate the Gaussian model-independent radii
in the core/halo model and compare the resulting ``model-independent"
 correlation functions to the prediction of the core-halo model.
Then we give a re-formulation of the ``model-independent" parameterization
of the correlation functions that can be applied to core/halo type of models
as well as systems without halo and clarify the model assumptions that
have to be made in order to achieve this ``model-independent" result.

In the followings, we use natural units, $\hbar = c = 1$, to simplify the notation.

\section{Correlation Functions and the $\omega$-Puzzle} \label{s:ch-BE}

Let us briefly recapitulate the theoretical description of the two-particle
Bose-Einstein correlation functions following the lines of 
refs.~\cite{chalo,nhalo}.

Particle emission is characterized by the single-particle 
Wigner-function $S(x,p)$. Here $x = (t,{\bf r}\,) $ denotes a point in
space-time and $p = (E, {\bf p} \,) $ a point in momentum-space.
The emitted particles are on mass-shell, $m = \sqrt{ E^2 - {\bf p}^{\, 2} }$.

An auxiliary quantity can be introduced as 
\bea
\tilde S(\Delta k , K ) & = & \int d^4 x \,\,
                 S(x,K) \, \exp(i \Delta k \cdot x ),\label{e:ax}
\eea
where
$
\Delta k  = p_1 - p_2$, $K  = {(p_1 + p_2)/ 2}$
and $\Delta k \cdot x $ denotes an inner product.

The invariant momentum distribution can be expressed as 
\bea
 E {\dst d n \ov d{\bf p} } & = & N_1({\bf p}) \,
= \, \tilde S(\Delta k = 0, K = p).\label{e:imd}
\eea
In the present paper, we utilize the so called hydrodynamical
normalization of the Wigner-functions,
\bea
\int
{\dst d{\bf p}\ov E} d^4x S(x,p) & = &
\int {\dst d{\bf p}\ov E} N_1({\bf p})  \,\, = \,\, \ave{n},\label{e:nimd}
\eea
where $\ave{n}$ is the mean multiplicity.

The two-particle correlation function can be written as
\bea
C_2({\bf p}_1,{\bf p}_2)
	& = & {\dst N_2({\bf p}_1,{\bf p}_2) 
		\ov N_1({\bf p}_1) \, N_1({\bf p}_2) }
	\, = \, 
	1 +  {\displaystyle\strut
         \mid \tilde S({\bf \Delta k} , {\bf K}) \mid^2 \ov
		\tilde S(0,{\bf p}_1) \tilde S(0,{\bf p}_2) }
	\, \simeq \,
	1 +  {\displaystyle\strut
         \mid \tilde S({\bf \Delta k} , {\bf K}) \mid^2 
	\ov \mid \tilde S(0,{\bf K})\mid^2 },
	\label{e:c2d}
\eea
e.g. in ref. \cite{pratt_csorgo,zajc}. 
Interestingly, this formalism can be introduced not only
when multi-particle symmetrizations are negligible, but even when 
they are fully taken into account~\cite{jzcst}.
For the most recent derivation,
where full multi-particle symmetrization effects were included, see ref.
~\cite{jzcst}.

The last approximation  in eq.~(\ref{e:c2d})
can be estimated to give correct results within 
5 \% error ~\cite{uli_l}.
We have neglected here final state interactions, 
and a completely chaotic particle  emission is assumed.

The second and the  higher order Bose-Einstein correlation functions,
\bea
C_n({\bf p}_1,{\bf p}_2,...,{\bf p}_n) & = &
	{\dst N_n({\bf p}_1,{\bf p}_2,...,{\bf p}_n) \ov
	N_1({\bf p}_1) N_1({\bf p}_2) ... N_1({\bf p}_n) }
\eea
are given~\cite{rrr} in terms of the Fourier-transformed Wigner-functions as
\bea
C_n({\bf p}_1,{\bf p}_2,...,{\bf p}_n) & = &
	{\dst \sum_{\sigma^n} \prod_{i=1}^n \tilde S(i,\sigma_i)
	\ov \prod_{i=1}^n \tilde S(i,i)} \,\, = \,\,
	\sum_{\sigma^n}  \prod_{i=1}^n {\dst  \tilde S(i,\sigma_i)
	\ov \tilde S(i,i)}  \,\, = \,\,
	\sum_{\sigma^n}  \prod_{i=1}^n  \tilde s(i,\sigma_i),
\eea
where $\sigma^n$ stands for the set of permutations of $n$ indexes,
\bea
\tilde S(i,\sigma_i) \, = \,
	 \tilde S({\bf K}_{i,\sigma_i},{\bf \Delta k}_{i,\sigma_i}),
	& \mbox{\rm and} &
	\tilde s(i,\sigma_i) \, = \,
	{\dst \tilde S(i,\sigma_i)
	\ov  \tilde S(i,i) }, \\
	{\bf K}_{i,\sigma_i} \, = \, 
		{\dst {\bf p}_i + {\bf p}_{\sigma_i} \ov 2},
	& \mbox{\rm and} &
	{\bf \Delta k}_{i,\sigma_i} \, = \, {\bf p}_i -  {\bf p}_{\sigma_i}.
\eea
Note the difference between $\sigma^n$, which stands for a set of 
permutations and $\sigma_i$ (subscript $i$),
which stands for the permuted value of index $i$ in a given permutation from
the set $\sigma^n$. 

Let us recapitulate the Assumptions of the core/halo model from ref.
~\cite{nhalo}:

\underline{\it Assumption 0:}
The emission function does not have a no-scale, power-law like structure.
This possibility was discussed and related to intermittency in
ref.~\cite{bialas}.

\underline{\it Assumption 1:} The bosons are emitted either from a {\it
central} part or
from the surrounding  {\it halo}. Their emission functions
are indicated by $S_c(x,p)$ and $S_h(x,p)$, respectively.
According to this assumption, the complete emission function can be written
 as
\bea
 S(x,p) = S_c(x,p) +  S_h(x,p),
\eea
using the hydrodynamic normalization of the Wigner functions.

\underline{\it Assumption 2:} We assume that the emission function which
characterizes
the halo changes on a scale $R_H$ which is larger
 than  $R_{max}\approx \hbar / Q_{min}$, the maximum length-scale
resolvable~\cite{chalo}
by the intensity interferometry microscope. However, the smaller central part
of size $R_c$ is
assumed to be resolvable,
 $ R_H > R_{max} > R_c $.
This inequality is assumed to be satisfied by all characteristic scales
in the halo and in the central part, e.g. in case the side, out or longitudinal
components~\cite{bertsch,lcms} of the correlation function are not identical.

\underline{\it Assumption 3:}  The momentum-dependent core fraction
$f_c(i) = N_c({\bf p}_i)/N_1({\bf p}_i)$ varies slowly on the relative
momentum scale given by the correlator of the core $\ts(1,2) \ts(2,1)$.

Let us also recapitulate the normalization conditions:
\bea
\int d^4x {\dst d{\bf p}\ov E} S_c(x,p) \, = \, \ave{n}_c,
& \mbox{\rm and}&
\int d^4x {\dst d{\bf p}\ov E} S_h(x,p) \, = \, \ave{n}_h,
\eea
where the subscripts $c,h$ refer to the contribution from
the central core  and from the halo, respectively.

Note that ref~\cite{chalo} also utilized  the above Assumptions,
however, in a modified form, that was based on 
Wigner-functions normalized to 1, and a core-fraction
$f_c = \ave{n}_c / \ave{n}$ was introduced, while in the present paper
a {\it momentum-dependent} core fraction 
$f_c({\bf p}) = N_c({\bf p}) / N({\bf p})$ is utilized.
 One finds that
\bea
 N_1({\bf p}) \, =  \, N_c({\bf p}) + N_h({\bf p}),
& \mbox{\rm and}&
\ave{n} \, = \, \ave{n}_c + \ave{n}_h.
\eea
According to ref. ~\cite{nhalo}, 
the general expression for $C_n(1,...,n)$ reads in the core/halo model as  
\bea
C_n(1,...,n) & = & 1 + \sum_{j = 2}^n
	\sum_{i_1, ..., i_j = 1}^{\null \,\,\, n \,\,\, _{\prime}}
	\sum_{\rho^n} \prod_{k=1}^j
	f_c(i_k) \tilde s_c(i_k,i_{\rho_k}).
	\label{e:fmix}
\eea
where $\sum'$ refers to  a summation over different 
values of the indices 
(i.e. $i_j = i_k$ type of terms are excluded). 
For the two-particle correlation function, the above equation takes a 
particularly simple form:
\bea
	C_2(1,2)  & = & 1 + f_c(1) f_c(2) \ts(1,2) \ts(2,1).
\eea
With the help of Assumption 3,
the core/halo model thus predicts the following form
for the two-particle correlation function:
\bea
{ C({\bf \D k}_{12}, {\bf K}_{12}) }
	 & = & 1 +
      \lambda_*({\bf K}_{12}) 
	{\dst \mid \tilde S_c( {\bf \D k}_{12}, {\bf K}_{12}) \mid^2 
		\ov \tilde S_c( 0, {\bf p}_1) \tilde S_c(0,{\bf p}_2)},
	 \,  \simeq \, 1 +
      \lambda_*({\bf K}_{12}) 
		{\dst \mid \tilde 
		S_c( {\bf \D k}_{12}, {\bf K}_{12}) \mid^2 \ov
                            | \tilde S_c( 0, {\bf K}_{12})|^2}, \nonumber \\
                           \label{e:lamq}
\eea
where one introduces an {\it effective intercept parameter} 
as 
\be
	\lambda_*({\bf K})  =
		\left[N_c({\bf K}) / N_1({\bf K}) \right]^2.
\ee
As emphasized in Ref.~\cite{chalo}, this {\it effective} intercept parameter
( which is not the same as the {\it exact intercept parameter, $\lambda_x = 1$} at $Q = 0$ MeV)
shall in general depend on the
mean momentum of the observed boson pair,
which within the errors of $Q_{min}$ coincides
with any of the on-shell four-momentum $p_1$ or $p_2$.

Thus one obtains the core/halo interpretation of the two-particle
correlation function:

The measured part of the BECF picks up an effective, momentum
dependent intercept parameter $\lambda_*({\bf K})$, 
that depends on the mean momentum only and which can be used as 
a tool to measure the momentum dependence of the fraction of particles
that are emitted from the core. 
On the other hand, the relative momentum dependence
of the BECF-s carries the information on the core  in this picture. 
Note, that in high energy heavy ion collisions the momentum
dependence of the $\lambda_*({\bf K})$ parameter is very weak, actually,
within the errors $\lambda_*({\bf K})$ is constant for the NA44 data
analyzed in ref.~\cite{chalo}. However, the validity of 
{\it Assumption 3} has to be checked experimentally for each
data set, by determining the momentum dependence of the
$\lambda({\bf K})$ parameter of the two-particle correlation function.

At this point,  we emphasize that
 non-Gaussian correlation functions with 
$0 \le \lambda_* \le 1$ are very well possible within the
core/halo picture, as discussed e.g. in refs.~\cite{chalo,3d}. 
We shall discuss in Appendix A under what conditions can the 
core correlator have an approximate Gaussian shape.

Since the core-halo model neglects possible partial coherence,
motivated by the success of fully chaotic Monte-Carlo simulations of
high energy heavy ion collisions like RQMD~\cite{RQMD}, the exact intercept
parameter is $\lambda_x = 1$ and the value of the exact intercept 
of the two-particle correlation function is always
 $1 + \lambda_x = 2$ in the core/halo picture. 
The effect of long-lived resonances is to create an unresolvable,
narrow peak in the region $0 < Q < Q_{min}$ from the interference
of the particles of $(h,h)$ and $(h,c)$ type, i.e. from those pairs
which have at least one member of the pair from the core. 
 This is illustrated in Figure ~\ref{f:shaded}, taken from Ref.~\cite{chalo}. 

%
%

Another very important general observation is that the region with small
$Q$ has a different characteristic structure from that of $Q > Q_{min}$
region. Thus any analytical expansions of the Bose-Einstein
correlation functions around the $Q = 0$ point become
both qualitatively and quantitatively unreliable for the core-halo
type of systems, characterized by $0 < \lambda_* < 1$. 
This point is further elaborated in the Appendix, let us return now to
the detailed discussion of the core-halo model using an example.

Note, that in principle the core as well as the halo part of
the emission function could be decomposed into more detailed contributions.
We shall argue below, that in case of
$S + Pb$ reactions at 200 AGeV at CERN SPS, {\it in the  NA44 acceptance},
one can separate the contribution of various long-lived resonances as
\bea
	S_{h}(x,p) \, = \, \sum_{r=\omega,\eta,\eta^{\prime},K^0_S}
		S^{(r)}_{h}(x,p)
		& \mbox{\rm and} &
	N_h({\bf p}) \, = \,\sum_{r=\omega,\eta,\eta^{\prime},K^0_S}
		N^{(r)}_{h}({\bf p}).
\eea

For a general consideration, e.g. the ones discussed
in ref.~\cite{nhalo}, the question whether the
$\omega$ decay products contribute to the halo or not, is
essentially indifferent.
However, when a data analysis is performed, this question
becomes important both qualitatively and quantitatively,
influence of the $\omega$ meson on the shape and the intercept
parameters of the two-particle correlation functions is suggested 
in various theoretical papers, see e.g. 
refs.~\cite{lcms,RQMD,res1,res2,restable,res3,uli-summ,uli-res}.

Let us briefly summarize the various statements made in this
scientific debate. Essentially, three different type of statements were
made regarding the influence of decay products of the $\omega$ meson on the  
short range part of the two-particle correlation function.
 Although we cannot repeat the fine-prints and all the reservations
that were made in earlier papers on this point, we think that 
a rough (and probably incomplete) summary can be made as follows:

\begin{description}
\item[{\bf A}] The effect of the $\omega$ decay products is {\it clearly} visible
on the two-particle BECF. These decay products are mainly responsible for the 
deviation of the BECF from a Gaussian shape and they influence the intercept
parameter also in a non-trivial manner. Essentially, the $\omega$ decays
contribute to the resolvable, core part of the source.
\item[{\bf B}]
	The decay products of the $\omega$ meson should be considered as
an {\it intermediate } case. Their contribution should be partially visible,
resulting in a deviation from the Gaussian shape of the BECF.
\item[{\bf C}]
	At the current level of experimental techniques, $Q_{min} \approx 
	5 - 10 $ MeV and $B(Q)_{max} \ge 25$ MeV, the $\omega $ decay products
	can be taken as part of the halo. Their contribution to the 
	two-particle BECF {\it cannot be resolved} experimentally in case of
	NA44 data.
\end{description}

	In the core/halo picture, the points {\bf A } - {\bf C} 
	correspond to the
	following conclusions: In case of {\bf A}, the $\omega$ decay
	products belong to the core.
	In case of {\bf B}, the core/halo
	model is not applicable to simplify the theoretical treatment
	of the two-particle BECF-s. This case corresponds to a
	really complicated situation.
	Finally, in case of {\bf C}, the $\omega$ decay products contribute
	to the halo and the theoretical description of the
	correlation function can be simplified substantially. 

	Let us also give a (probably incomplete) list of papers where
	these statements were made. Conclusion {\bf A} has been reached
	in ref.~\cite{lcms}, in a Monte-Carlo simulation
	with the help of SPACER, and later in ref.~\cite{res3} assuming $5 $
	MeV $Q_{min}$ resolution for $Pb + Pb$ reactions at CERN SPS in
	a 3d hydrodynamical simulation with the code HYLANDER.
	Conclusion {\bf B} was reached 
	in refs.~\cite{restable,uli-summ,uli-res,sinyu96}, 
	using phenomenological parameterization of a hydrodynamically evolving
	core with thermally populated resonance production.
	In all these studies, full resonance decay kinematics was included
	and resonance production was described either in a thermal manner
	or in a non-equilibrium process with the help of re-scattering.
	All of these studies share in common that they evaluated
	the correlation function as a mathematical function
	but they did not consider in detail the error bar distribution
	on the theoretical curve.

	Finally, in refs.~\cite{chalo,nhalo,simon,csorgo-kiang},
	conclusion {\bf C} was reached based on considerations related
	to the NA44 detector resolution and the surprisingly Gaussian
	shape of the measured NA44 correlation functions.

	Another interesting study was performed recently by Padula and 
	Gyulassy~\cite{gyp96}, that included a simulation of binning and detector
	resolution effects for AGS energies with the help of the 
	inside-outside cascade code CERES, with proper resonance
	decay kinematics, to conclude that 

\begin{description}
\item{\bf D} 
	Preliminary data seem to rule out dynamical models with
	significant $\omega$ and $\eta$ resonance fraction yields
	in case of the E802 measurement of $Si + Au$ data at 14.6 AGeV
	at Brookhaven AGS. 
\end{description} 

	We do restrict our study to CERN SPS reactions,
	and published NA44 data. In the next section, we perform a Monte-Carlo
	simulation to make a connection between source parameters and 
	the parameters of the fitted correlation functions 
	before attempting the data analysis.

\section{Monte Carlo Simulations}\label{s:ch-MC}

For the sake of simplicity, one may simplify the problem by fixing the
mean momentum of the observed particle pairs and introduce the 
source density for those particles which are emitted with the given mean momentum
as
$\rho_{\bf K}(x) = S(x,\bk)$. 
The index ${\bf K}$ will thus be suppressed in the forthcoming,
but implicitly we shall assume that the analysis is performed at a 
fixed value of $\bk$. Also, in the applications, we shall analyze 
NA44 data where the mean momentum of the pairs is restricted to 
a certain range.
We have had access only to the one dimensional slices of 
a three-dimensional NA44 data set,
along the main axis in the LCMS frame~\cite{lcms}. 
These data were taken at a fixed mean momentum. 
In effect, we thus analyzed three different one dimensional data set instead
of analyzing a single three dimensional one.
As the quality of the published data is expected to be improved in the 
near future, we hope that this limitation
will disappear and in a future analysis one may study the 3d correlation
functions at a fixed value of ${\bf K}$ directly in a similar manner 
as presented below. 

The limitation that we had access to 1d slices of data yields an
advantage as well, since it is enough  for the present paper
to formulate the procedure for one dimensional distributions only.
The generalization of the method for multi-dimensional distributions
is trivial hence it will be omitted.

In the core/halo picture, the source density is written as
\begin{equation}
\rho(x)=f_c\rho_c(x)+(1-f_c)\rho_h(x)
\end{equation}
where $f_c$ from now on represents the fraction of pions produced in the core
at the given mean momentum $\bk$. 
In order to make
predictions for the model, forms for $\rho_c(x)$ and $\rho_h(x)$ must be
assumed. With Gaussian assumptions
for both, $\rho_i(x)=\frac{1}{\sqrt{2\pi R_i}}e^{-x^2/2R_i^2}$, the resulting
correlation function is
\begin{equation}
\label{e:chalo}
C_2(Q)=1+f_c^2 e^{-R_c^2 Q^2}+(1-f_c)^2e^{-R_h^2
Q^2}+2f_c(1-f_c)e^{-\frac{1}{2}(R_c^2+R_h^2)Q^2}
\end{equation}

%
%

Examination of this expression reveals Gaussian terms for the core, the halo,
and an `interference' term. Figure \ref{terms} shows the relative size of these
terms, as a function of some momentum variable $Q$, for typical values of
$f_c$, $R_c$, and $R_h$.
 The effect of the halo is to introduce a sharp peak in the correlation
function at low values of $Q$. This narrow peak in $Q$-space is indicative of a
large length-scale in $x$-space. Note that the model predicts $C_2(Q=0)=2$ --
there is no need to assume any coherent boson production which would reduce the intercept. Figure
\ref{fc} illustrates the effect of the $f_c$ parameter on the correlation
function.

%
%

Since we are more interested in the radius parameters of the core 
than the halo, the
easiest way to extract the proper core parameters is to remove the low $Q$ data
points so that the effect of the halo is negligible,
a procedure suggested e.g. in ref.~\cite{chalo}. 
This simple method minimizes the effect
of the Gaussian assumption for the shape of the halo structure - as 
long as the length-scale of the halo is large compared to $\hbar/Q_{min}$,
 removing the low $Q$ points should remove the halo's
effects. Fitting the remaining data with a standard Gaussian form 
should produce $R=R_c$ and $\lambda=f_c ^2$.

However, given the finite experimental resolution of $\sim$10 MeV/c, the halo
may not be visible at all. If this is the case, the extracted $\lambda$ and $R$
parameters still have an interpretation as the fraction of core pions (squared)
and the core radius parameter, respectively. 
If they do not change as $Q_{min}$, the minimum value of $Q$
included in the data, is increased, it will indicate that the correlation
function really is Gaussian. This would contradict the prediction that some
resonances distort the Gaussian shape of the correlation function.

In either case, we must establish the limits of the procedure before attempting
the analysis of real data. 
There must be some maximum number of data points that can be
removed from a real correlation function while still being able to extract
valid parameters. If that
 number is too small, then the idea of removing low $Q$ points in order to
study the correlation function  of the core is invalid - 
if the length-scales of the core and the halo cannot be separated, 
then the fitted parameters of the truncated correlation function
become dependent on $Q_{min}$, the size of the excluded region.

The core-halo model takes advantage of finite experimental resolution -- as
resolution increases, and more of the low $Q$ structure of $C_2(Q)$ is
revealed, perhaps a more detailed model will be necessary. But for now, this
reasonable model imitates the effects of resonance production and can 
attribute physical significance to the $R$ and $\lambda$
parameters extracted from a correlation function.

\subsection{The acceptance-rejection method}
What is required is a method of producing correlation functions where the
underlying parameters are known in advance, so they can be compared to the
extracted parameters. This will be done by simulating the 
actual distribution $A(Q)$ and  the background distribution $B(Q)$
that a real experiment 
uses to produce $C_2(Q) = A(Q) / B(Q)$.

It is necessary to assume a form for $A(Q)$ and $B(Q)$ in order to do this. The
background distribution $B(Q)$ is different for every experiment, so one must
parameterize it appropriately each time. The $A(Q)$ distribution will not be
parameterized directly;
 rather, we will assume a form for $C_2(Q)$, and then simulate the
distribution $A(Q)=B(Q) \cdot C_2(Q)$. In this way, we can have control over
the parameters that are going into the simulation.
The finite number of iterations performed in the simulation ensure that these
parent distributions are never reproduced exactly, so the assumed form for
$C_2(Q)$ is not returned {\it directly}. Binning effects are automatically
included and, most importantly, the {\it error-bar distribution} on
the correlation function will 
follow the experimental one at least when statistical errors are considered. 
Fits shall be done with optimizing $\chi^2/NDF$ hence the error bar distribution
plays an important role when determining the best fitted values and
their errors.

We would like to emphasize that the simulation of the $A(Q)$ and $B(Q)$
distributions is {\it essential} when discussing possible resonance effects
on the two-particle correlation function, since the resonance decay 
products are expected to influence the shape of the Bose-Einstein
correlation function in the $Q \le 50$ MeV 
region~\cite{lcms,res2,chalo,uli-summ,uli-res}
and they also contribute to the fitted value of the radius and the
intercept parameters. However, the fitted values are strongly 
influenced by the distribution of the errors on the data points,
since the fit selects to reproduce the best
that part of the correlation function,
which has the smallest errors. Unfortunately, very few experiments
decided to publish their actual and background distributions,
but some~\cite{afs,na44hbt1,na44hbt2,lowpt,highpt} did this important step.
Even fewer theoretical models tried to reproduce the distribution
of the statistical errors on the data as arising from the number
of actual and background pairs as a function of the relative momentum.
In fact, none of the theoretical simulations that reached a
conclusion belonging to type {\bf A} or {\bf B} included this
important step.

%
%

Unfortunately, it is not possible to sample from 
the complicated distributions $A(Q)$ and $B(Q)$
directly. Instead, we will use the acceptance-rejection method,
 which works as follows:
Let the known distribution to be sampled be denoted by $p(Q)$,
that will be chosen as $A(Q) $ or $B(Q)$ in the subsequent parts. 
The probability distribution $p(Q)$ should be
normalized, $\int p(Q)\,dQ=1$, over the range of interest. Find a comparison
function, $f(Q)$, which has the
 following properties:
\begin{enumerate}
\item Let $f(Q)$ satisfy $f(Q)>p(Q)$ for all $Q$.
\item There exists a closed form for a random $Q$ point sampled from $f(Q)$.
\end{enumerate}
In the interest of computational efficiency, the comparison function $f(Q)$
should be chosen to be as close to $p(Q)$ as possible, while still satisfying
$f(Q)>p(Q)$, in order to minimize the number of rejected points.
A random $Q$ point, $Q_i$, is sampled from the comparison function $f(Q)$. Then
the value of the parent distribution $p(Q_i)$ is calculated. The ratio
$R=p(Q_i)/f(Q_i)$, which is always less than 1, is then compared to another
uniform random number $U'$.
If $U' > R$, the
point $x_i$ is rejected and the process starts over again. If $U' <R$, $Q_i$ is
accepted. In this way, the distribution $p(Q)$ is sampled.

\subsection{Simulating a one dimensional correlation function}

If we assume the halo form for $C_2(Q)$, eq. \ref{e:chalo}, the two
distributions to be simulated are $B(Q)$ and $A(Q)$.
For the CERN experiment NA44, studying 
S+Pb collisions, the shape of $B(Q)$ is given approximately by \cite{fax}
\be
B(Q)  =  Q^3 e^{-3.6Q^{0.3}},
\ee
hence the $A(Q)$ distribution is sampled as
\be 
A(Q)  =  B(Q)(1+f_c^2 e^{-R_c^2
Q^2}+(1-f_c)^2e^{-R_h^2 Q^2}+2f_c(1-f_c)e^{-\frac{1}{2}(R_c^2+R_h^2)Q^2}).
\ee

%
%

For these one-dimensional simulations, $Q$ can be any relative momentum variable
-- $Q_{inv}$, $Q_{side}$, $Q_{out}$, etc. The above expression for $B(Q)$ was
given for $Q_{inv}$, but it is valid for any one dimension of $\mathbf{Q}$ as
long as the other components are small.
The parameters for a Lorentzian comparison function are adjusted by hand
until the Lorentzian satisfies condition 1) reasonably well. 
The Lorentzian comparison function was sampled as 
suggested in ref.~\cite{numrec}. With the help of Lorentzian comparison
functions, the distributions $A(Q)$ and $B(Q)$ 
were sampled as described above. 
As each point was sampled from either $A(Q)$ or $B(Q)$, it is binned. 
The bin size of 10 MeV/c is chosen to
reflect experimental situation for the multi-dimensional
NA44 data analysis. 
The value of $Q$ for each bin is calculated by averaging all
entries, from both $A(Q)$ and $B(Q)$, that fall in that bin.
Figure \ref{mc1} shows the results of the simulations of $A(Q)$, $B(Q)$, and
the resulting $C_2(Q)$. The number of iterations for the simulation is chosen
to give reasonable error bars on the final $C_2(Q)$. The number of iterations
must be higher for $A(Q)$, however, since $A(Q)>B(Q)$ for all $Q$. The areas under the curves are
used to calculate the proper ratio of iterations.

The error bars are calculated as follows. For $A(Q)$ and $B(Q)$, we assume that
the number of entries in each bin is Poisson distributed, so that the error for
each bin is just the square root of the number of bin entries. The errors for
$C_2(Q)$ are then
 just combined in the usual manner.

Now we can fit the simulated correlation function with a standard Gaussian
shape, $C_2(Q)=1+\lambda e^{-R^2 Q^2}$, reflecting experimental analysis
techniques, and observe the behavior of the parameters as
$Q_{min}$ is increased. Figure \ref{mc2} shows the value of the $R_*$ and
$\lambda_*$ parameters as low $Q$ data points are removed. The solid lines
represent the input values, $R_c=4.0$ fm and $f_c^2=0.5625$.
Examining these results indicates that reliable values for the parameters can
be extracted, i.e. $R_* = R_c$ within errors and
$\lambda_*  = f_c^2$ within errors,
 when $Q_{min}$ is varied in the range of 10 to 50 MeV/c. We can also
read off from Figure ~\ref{mc2} that the 
fitted radius  and intercept values are essentially 
insensitive to the exact value of $Q_{min}$ if this is varied in 
the above range.

\section{Core/halo model analysis of NA44 data}\label{s:ch-anal}
We now apply the data chopping analysis, tested in the 
earlier section, to real experimental data. CERN experiment NA44 has
been running since 1991, and has an extensive collection of correlation
function data \cite{na44hbt1,na44hbt2,na44npa,dodd,lowpt,highpt}. 
We analyze their S+Pb data. The data set is three-dimensional --
that is, it covers all of the (out, side, long) space, but we only have access
to slices of the data along the three axes. So in effect we are analyzing three
one-dimensional slices of the three-dimensional Bose-Einstein correlation 
functions, for each published data set. 
Values of the two-particle correlation functions and their errors
 were read directly off the published figures by eye,
as accurately as possible.
 NA44 has analyzed S+Pb data for both $\pi^+$ and
$K^+$. The $\pi^+$ data were taken with two different magnetic settings  
-- `high $p_t$' data is characterized by $<p_t>=450
MeV/c$, and the `low $p_t$' data has $<p_t>=150 MeV/c$. This  was done in an
attempt to observe some detailed dynamics of the system.

Thus we have analyzed the slices of the three-dimensional NA44
Bose-Einstein correlation function for the low-$p_t$ and high-$p_t$ pion
sample and for the kaon sample in $S+Pb$ reactions at CERN SPS, data were
from refs.~\cite{highpt,kaon}.
Unfortunately, we had no access to the three-dimensional distributions
either for NA44 measurements or for other data. Ideally, the analysis
presented below should have been performed on the (unpublished)
three-dimensional distributions. However, our final result indicates that
the extracted values for the intercept parameters are within errors similar
for each projection of the NA44 sample, which justifies 
the utilization of the projections in this particular case.

Our analysis results are shown in the Figures ~\ref{f:lpr}-\ref{f:ki}. 
Note that the
left-most data point in each figure represents the value of that particular
parameter when \emph{none} of the data points are removed. 
Also, with three separate slices along the three main axis 
instead of one three-dimensional set,
 one can extract three values for the $\lambda$
parameter. The NA44 analysis produces just one $\lambda$, since they fit to
\begin{equation}
C_2(Q)=1+\lambda e^{-Q_{out}^2 R_{out}^2 -Q_{side}^2 R_{side}^2 -Q_{long}^2
R_{long}^2}
\end{equation}

%
%

%
%

%
%

%
%

%
%

%
%

The parameters show no significant, systematic change as $Q_{min}$ is
increased. This confirms that the shape of the correlation function really is
Gaussian. In the context of the core-halo model, this suggests that long-lived
resonances are not resolved by NA44, neither in case of pions, nor for
kaons, see Figures~(\ref{f:lpr}-\ref{f:ki}). 

In order to extract values for $R$ and $f_c$ from this analysis, we fit the
previous data with a constant (see Table ~\ref{results}). The data up to
$Q_{min} = 40 $ MeV/c  are chosen, since this has been shown to be a reliable
range.
\bigskip
\begin{table}[ht]
\centering
\begin{tabular}{|c|c|c|c|}
\hline\hline
parameter & high $p_t$ $\pi^+$ & low $p_t$ $\pi^+$ & $K^+$ \\ \hline
$R_{c \ out}$ (fm)& 2.92 $\pm$ .13 & 4.29 $\pm$ .13 & 2.54 $\pm$ .18 \\
$R_{c \ side}$ (fm)& 2.90 $\pm$ .18 & 4.24 $\pm$ .26 & 2.22 $\pm$ .19 \\
$R_{c \ long}$ (fm)& 3.31 $\pm$ .16 & 5.43 $\pm$ .30 & 2.67 $\pm$ .22 \\
$f_{c \ out}$ & .704 $\pm$ .012 & .725 $\pm$ .011 & .802 $\pm$ .027 \\
$f_{c \ side}$ & .735 $\pm$ .021 & .647 $\pm$ .021 & .736 $\pm$ .033 \\
$f_{c \ long}$ & .738 $\pm$ .014 & .724 $\pm$ .015 & .789 $\pm$ .029 \\
\hline\hline
\end{tabular}
\caption{Extracted values for $R_c$ and $f_c$ from NA44 S+Pb data.}
\label{results}
\end{table}

These results show several things.
As expected, the kaon core fraction is larger than either pion fraction.
 Further, the trend in the core radii is
$R_{kaon} < R_{high p_t} < R_{low p_t}$. The interpretation of this
result is not a trivial task~\cite{3d}. One would first imagine that
the decrease of these radii follows the trend of increasing
resonance influence, suggesting that short-lived resonances can also affect
core radii. It is also possible that kaons decouple from the 
interaction region sooner than pions do.
Secondly, the core fractions are similar for the high and low 
$p_t$ data, since the intercept parameter is within errors
independent of the transversal mass of the  particles.
One might argue that the constancy of the $\lambda$ parameter
 implies that any increase in resonance contribution
 in the low $p_t$ data takes place in the short-lived resonances. 

\bigskip
\begin{table}[ht]
\centering
\begin{tabular}{|c|c|c|c|c|c|}
\hline\hline
Name & $m$ (MeV) & $\tau$ & Decay & $f_{Fritiof}$ & $f_{RQMD}$ \\ \hline
direct $\pi$& 140 & 7.804 m & - & 0.19 & 0.33\\
$\rho$ & 770 & 1.3 fm & $\pi \pi$ & 0.40 & 0.26 \\
$\Delta$ & 1232 & 1.64 fm & $N\pi$ & 0.06 & 0.12 \\
$K^*$ & 892 & 3.94 fm & $K\pi$ & 0.09 & 0.07 \\
$\Sigma(1385)$ & 1385 & 5.5 fm & $\Sigma \pi$ & 0.01 & 0.02 \\ \hline
$\omega$ & 783 & 23.4 fm & $\pi \pi \pi$ & 0.16 & 0.07 \\
$\eta \prime$ & 958 & 982 fm& $\eta \pi \pi$ & 0.02 & 0.02 \\
$\eta$ & 548 & $1.64 \times 10^5$ fm & $\pi \pi \pi$ & 0.04 & 0.03 \\
$K_S^0$ & 498 & 2.7 cm & $\pi \pi$ & 0.03 & 0.07 \\
\hline\hline
\end{tabular}
\label{res}
{\small
\caption{Sources of pions in relativistic heavy ion collisions, taken
from Ref.~\cite{restable}.}
}
\end{table}

This interpretation, however, is not likely since
hydrodynamical expansion may also result in a strong transversal
mass dependence of the measured radius parameters~\cite{3d},
even an approximate $m_t$ scaling can be obtained in this manner
in a certain limiting case.
Further, a comparison of 
Tables~\ref{results} and ~\ref{res} indicates that the measured
core fractions are in the vicinity of the core fractions of 
Fritiof and RQMD if $\omega$, $\eta$ and $\eta^{\prime}$ are taken
as unresolved long lived resonances, as indicated by Table ~\ref{comp}. 

\bigskip
\begin{table}[ht]
\centering
\begin{tabular}{|c|c|c|c|}
\hline\hline
$f_{c,NA44}$ (high $p_t$ $\pi^+$ ) & $f_{c,NA44}$ (low $p_t$ $\pi^+$ ) 
& $f_{c,Fritiof}$ & $f_{c,RQMD}$ \\ \hline
	0.71 $\pm$ 0.01 & 0.71 $\pm$ 0.01 & 0.75	&	0.80 \\
\hline\hline
\end{tabular}
\caption{Core fractions $f_c$ from NA44 S+Pb data 
compared to core fractions from RQMD and Fritiof,
when $\omega$, $\eta$ and $\eta^{\prime}$ are taken as unresolved
long-lived resonances. The core fractions for the NA44 low $p_t$ and
high $p_t$ sample are obtained from a simultaneous fit to 
($f_{c,out},f_{c,side},f_{c,long}$) by a constant value, and the result
is rounded to two decimal digits. }
\label{comp}
\end{table}
 
Since both $\omega$ and $\eta$ decay predominantly 
into three pions, in a narrow region of phase space characterized by
$m - 3 m_{\pi}$, these resonance decays should produce predominantly
low-$p_t$ pions. Similar conclusion holds also for the
   $\eta^{\prime}$ resonance where the phase space is characterized by
$m_{\eta^{\prime}} - 2 m_{\pi} - m_{\eta}$, which is also rather small.
Thus a resonance interpretation of the $m_t$ dependence of the
radius parameters seems to contradict to  the 
observed constancy of the $\lambda$ parameters. Collective effects,
like three dimensional hydrodynamic expansion and $m_t$ dependent
volume factors that enhance the direct production of pions at low
$p_t$ as compared to the direct production of heavier
resonances were discussed as a possible solution of this
puzzle in Refs.~\cite{3d,chalo,vck}.

In the future, we hope that this type of analysis will be
applied to full three-dimensional data sets by various experimental groups.
 This would be more significant for several reasons. First, several
recent papers \cite{uli_s,cross2} have suggested the presence of cross-terms
in the 3-D correlation function, 
something which cannot be detected with just slices of data along the
three axes.  Secondly, the 1-D slices we have used may cover up some of the
low-$Q$ detail, since they integrate over small regions of the other two
dimensions (typically $<40$
 MeV/c). 

Looking farther ahead, the 
core-halo model could be useful in the next generation of
heavy ion experiments. Physicists at RHIC or at CERN experiment NA49
 expect to have a $Q$ resolution of
$\le$ 5 $MeV/c$, which could be sufficient to see the halo directly.

\section{Conclusions}\label{s:ch-concl}
In summary, an outline of Bose-Einstein correlation analysis for heavy ion
collisions has been presented. In contrast to theoretical predictions of a
complicated correlation function due to the effect of resonance decays,
experimental correlation functions
 seem to be characterized by just a single intercept parameter $\lambda$ 
and a set of Gaussian radius 
parameters, $R_{side}$, $R_{out}$, $R_{long}$ (and $R_{out-long}$).
 As an attempt to incorporate 
resonance effects into a simple correlation model, core
and resonance halo sources are dealt with separately, each characterized by a
Gaussian width parameter along with a core fraction parameter. 
 Data from NA44 do indicate indirectly 
the presence of such a halo.  Our results 
cast doubt on the validity of predictions that long-lived resonances
produce complicated structure
 in the Bose-Einstein correlation function at CERN SPS, when observed 
with a two-particle resolution of $Q_{min} \approx 10$ MeV.

In the Appendix,
we present one of the simplest possible
application of the core-halo model to illustrate that the Gaussian
``model independent" HBT radii are {\it unreliable qualitatively}
and are {\it unreliable quantitatively} in their frequently 
quoted original form, if the intercept parameter $\lambda < 1$.
At the end of the Appendix, we include the corrected definition
of these parameters, that can be applied to core/halo type of 
systems with $\lambda < 1 $ also.

\bigskip
\leftline{\large\bf Acknowledgments:} 
T. Cs\"org\H o would like to thank to Professors Kiang and Gyulassy
for their kind hospitality at Dalhousie and Columbia Universities.
This research was supported by the OTKA Grants W01015107, T016206,
T024094 and by an Advanced Research Award of the Fulbright Foundation
and by  an operating grant from the Natural Sciences and Engineering
Research Council of Canada.
\medskip
\vfill\eject
\bigskip
\leftline{\large\bf Appendix A: What do ``Model-Independent Radii" measure?}
\medskip

In this Appendix, we discuss the Gaussian model-independent radii
of the HBT correlation functions in terms of the core/halo model,
along the lines of ref.~\cite{tect}. 

Let us recall the definition of the ``model-independent" Gaussian
radius parameters:
\bea
	R_{i,j}^2 & = &
		 \langle (x_i - \beta_i t) (x_j - \beta_j t) \rangle
		 - \langle (x_i - \beta_i t)\rangle
		   \langle (x_j - \beta_j t) \rangle, \\
	C(\bk,\bdk ) & = & 1 + \exp\left(- R^2_{i,j} Q_i Q_j \right), \\
	N({\bf p})  & = & \langle 1 \rangle, \\
	\langle f(x,p) \rangle & = & \int d^4 x f(x,p)  S(x,p),
\eea
	where $i = side, out$ or $long$, $\beta_i = K_i/K_0$ is the
	component of the four-velocity vector of the pair in the
	direction $i$ and $Q_i = \bdk_i$, and $S(x,p)$ is
	the emission function that characterizes a chaotic 
	system, see ref.~\cite{uli_s,uli_l}.

	For clarity, it is better to consider a one-dimensional system only.
	Let us consider a core/halo type of system, with Gaussian
	ansatz for both the core and the halo part, with a core radius of
	$R_c = 4 $ fm, a halo radius of $R_H = 40 $ fm,
	and a core-fraction is assumed to be $f_c = 0.75$.
	Let us compare the full correlation function $C_2(Q)$,
	given by eq.~(\ref{e:chalo}) to the
	core/halo model approximation $C_2^{c/h}(Q)$, given as
\bea
	C_2^{c/h}(Q) & = & 1 + \lambda_* \exp( - R_*^2 Q^2) \\
	\lambda_* & = & f_c^2 \, = \, 0.56 \\
	R_* & = & R_c \, = \, 4 \mbox{\rm ~fm}
\eea
	and to the model-independent
	Gaussian approximation to correlation function, $C_2^{G}(Q)$,
	which reads as
\bea
	C_2^G(Q) & = &  1 + \lambda_G \exp( - R_G^2 Q^2), \\
	\lambda_G & = & 1 \quad \mbox{\rm by definition}, \\
	R_G^2 & = & \langle x^2 \rangle - \langle x \rangle^2
	\qquad \rightarrow \quad R_G = 20.3 \mbox{\rm ~fm}.
\eea
	This comparison is visualized in Figure ~\ref{f:append}.

%
%

	In Figure ~\ref{f:append} it is clearly visible,
	 that the model-independent
	Gaussian parameterization of the correlation function works
	only in the $ Q < \hbar / R_h = 5$ MeV region. The variance
	of the source function is dominated by the variance of the 
	halo part, and this is picked up by the model-independent
	Gaussian radii. The approximation works in the unobservable
	$Q$ range, however, in the well measurable $Q > 10 $ MeV region
	the resulting correlation function approximates rather incorrectly
	the full correlation function. On the other hand, the core/halo
	model approximation fails in the unobservable $Q < 5 $ MeV region,
	but it approximates with high accuracy the correlation function
	in the observable $Q > 10 $ MeV. If a measurement is performed
	in the $ Q > 10 $ MeV range, the extrapolated correlation function	
	will be a Gaussian with $\lambda  < 1$. We may conclude, that
	for a core/halo type of system, the Gaussian model-independent
	approximation yields qualitatively incorrect results in the range
	$Q > \hbar / \langle x \rangle^2_h$, when
	naively compared to a measurement of such a correlation
	function, since $\lambda < 1$ will be measured experimentally. 
	If the radius parameter is considered only, its value is 
	quantitatively wrong ($R_G = 20.3$ fm instead of 4 fm)
	 in the model-independent approximation~\cite{uli_s}, 
	since $R_G^2 = \langle x^2 \rangle - \langle x\rangle^2$
	is dominated by the contribution of the halo. 

	Note that the authors of refs.~\cite{uli_l,uli_s} warned the readers
	against the straightforward  applications of their result. Let
	us quote their warning from ref.~\cite{uli_l}:
	{\it `` We would like to point out, that one must take care when
	comparing the above radii with the experimentally measured
	correlation radii, since the former measure second derivatives
	around $Q = 0$, while the latter are parameters of a
	Gaussian fit to the whole correlation function ..."}.

Such non-Gaussian structures appear
e.g. in the case of stable distributions, like the Lorentzian distribution,
for which the Fourier-transformed distribution exists, however the 
first or the second moment of the distribution diverges thus the 
Fourier-transformed distribution is not analytic at small values of the
relative momenta. See appendix of Ref.~\cite{3d} for further details 
on this point. Such a non-Gaussian correlation function may characterize
even the core part of the distribution.
	
\begin{figure}[t]
%
%
\begin{center}
	\leavevmode\epsfysize=3.5in
\phantom{MMMM}          
          {\epsfbox{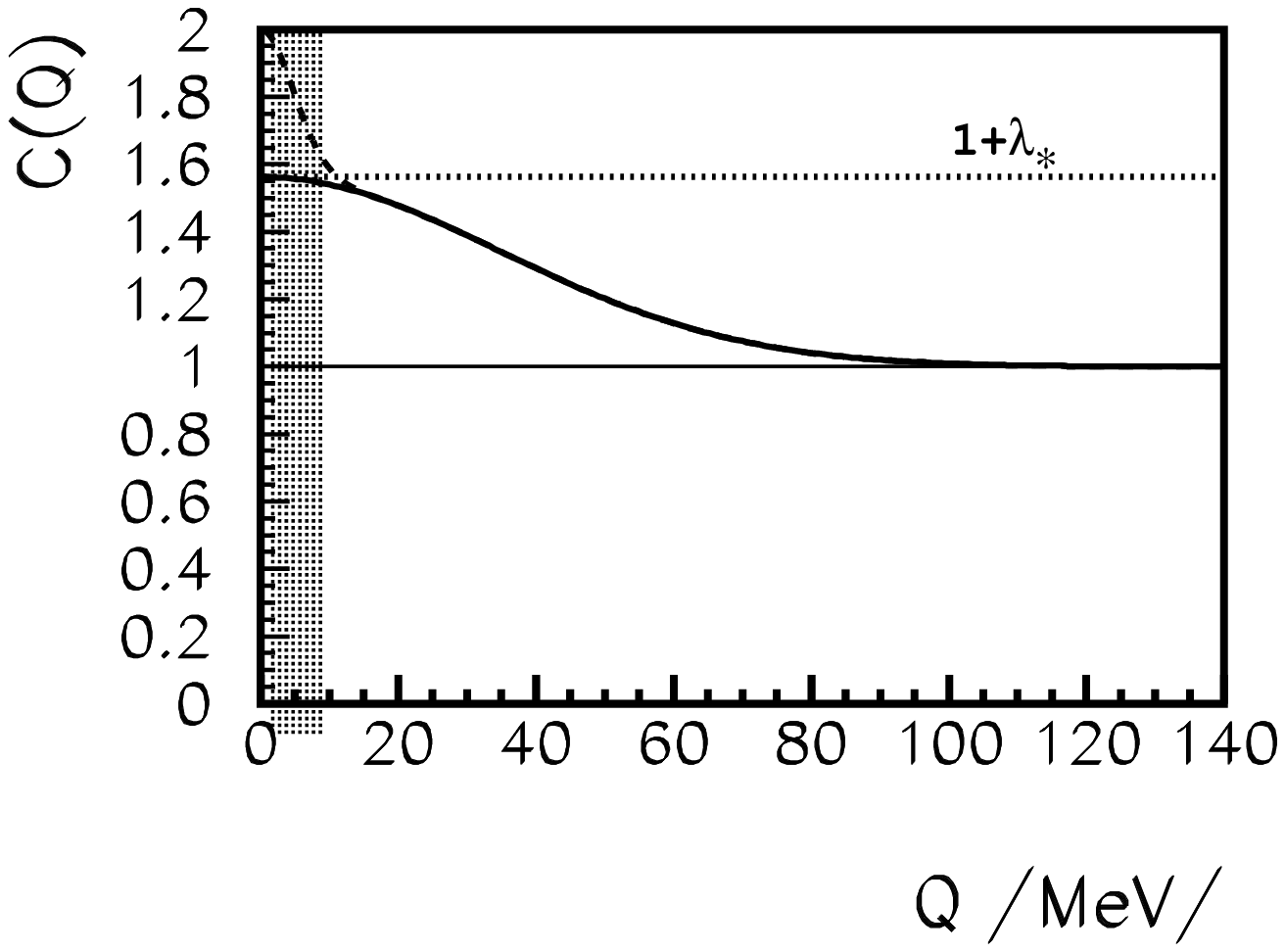}}
\end{center}
\caption{
}
\label{f:shaded}
\end{figure}

\begin{figure}[t]
%
%
          \leavevmode\epsfysize=3.3in
          \epsfbox{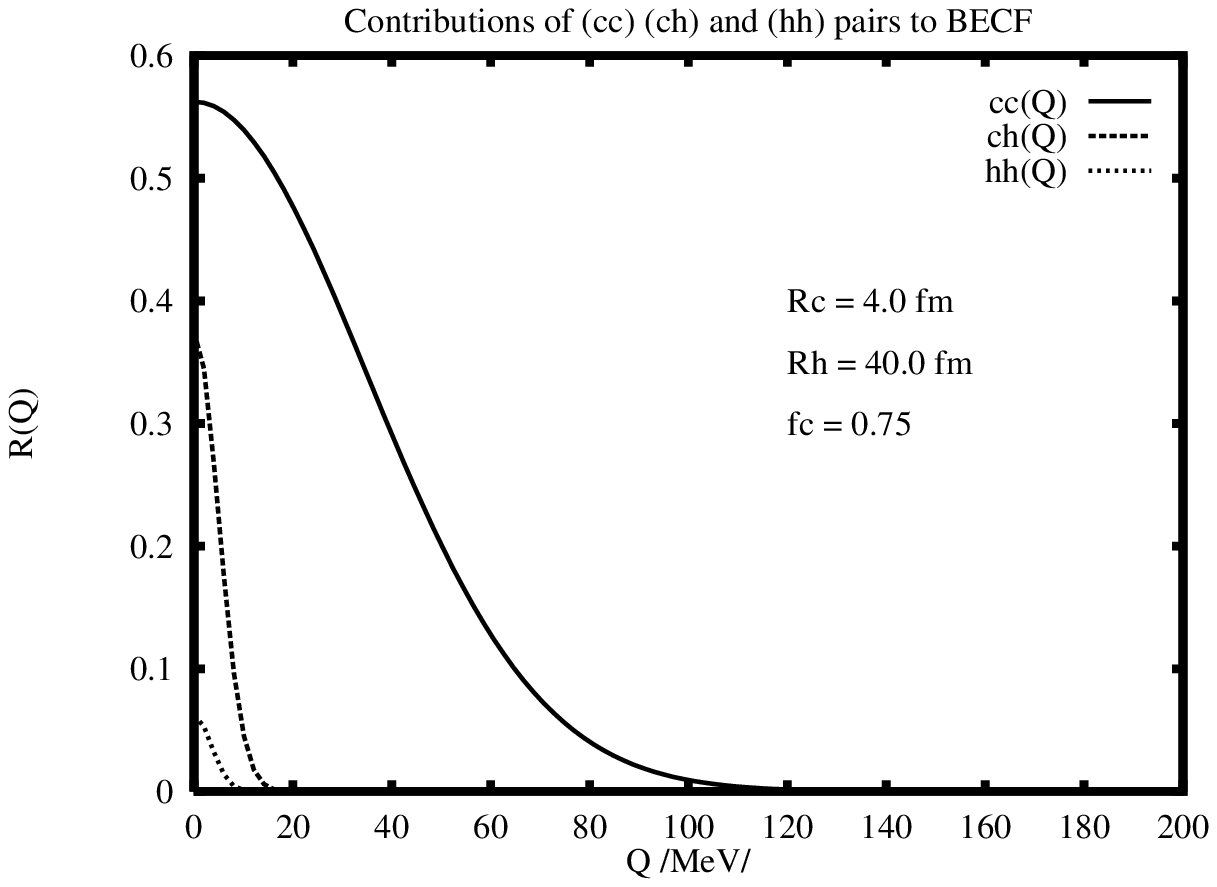}
\caption{
}
\label{terms}
\end{figure}

\begin{figure}[t]
%
%
          \leavevmode\epsfysize=3.3in
          \epsfbox{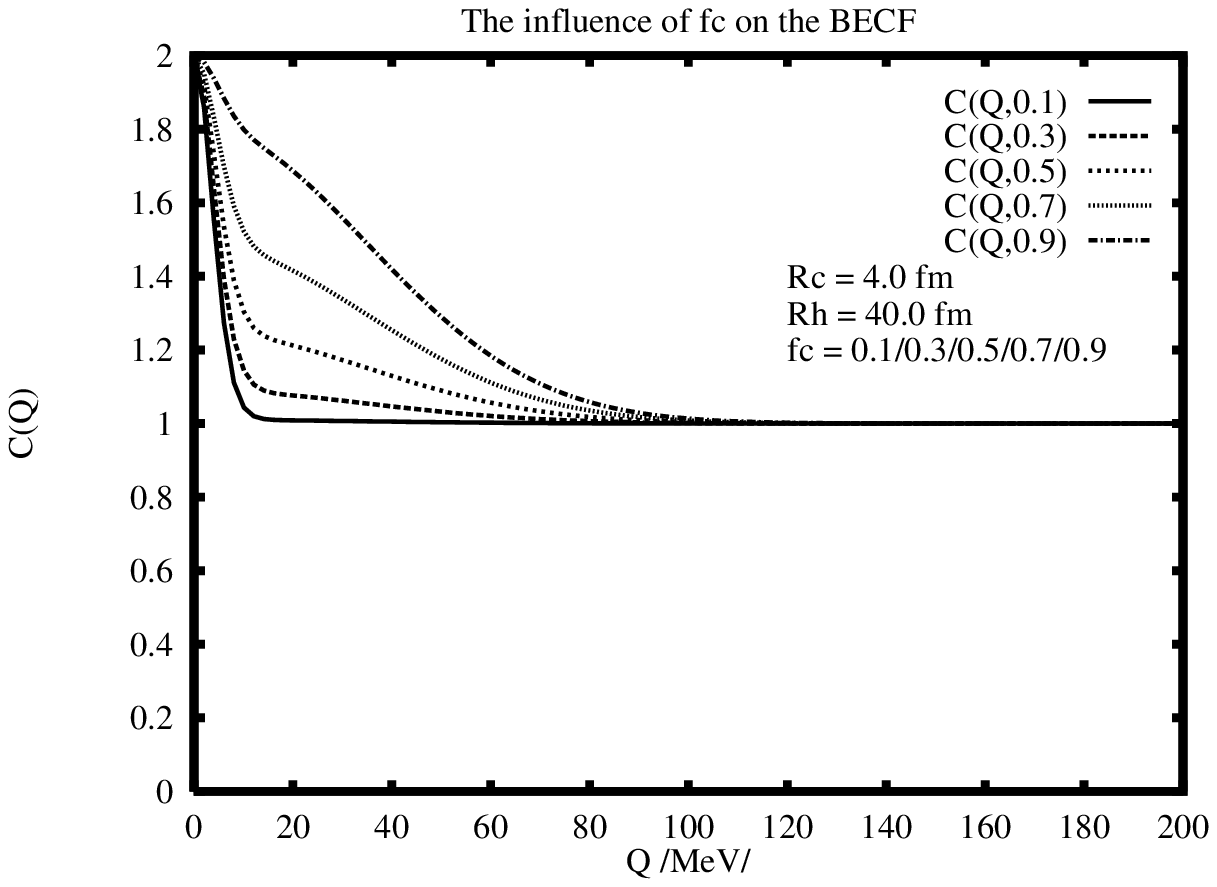}
\caption{
	}
\label{fc}
\end{figure}

\begin{figure}[t]
%
%
\begin{center}
\vskip -2cm
\hskip -2 cm
          \leavevmode\epsfysize=9.0in
          {\epsfbox{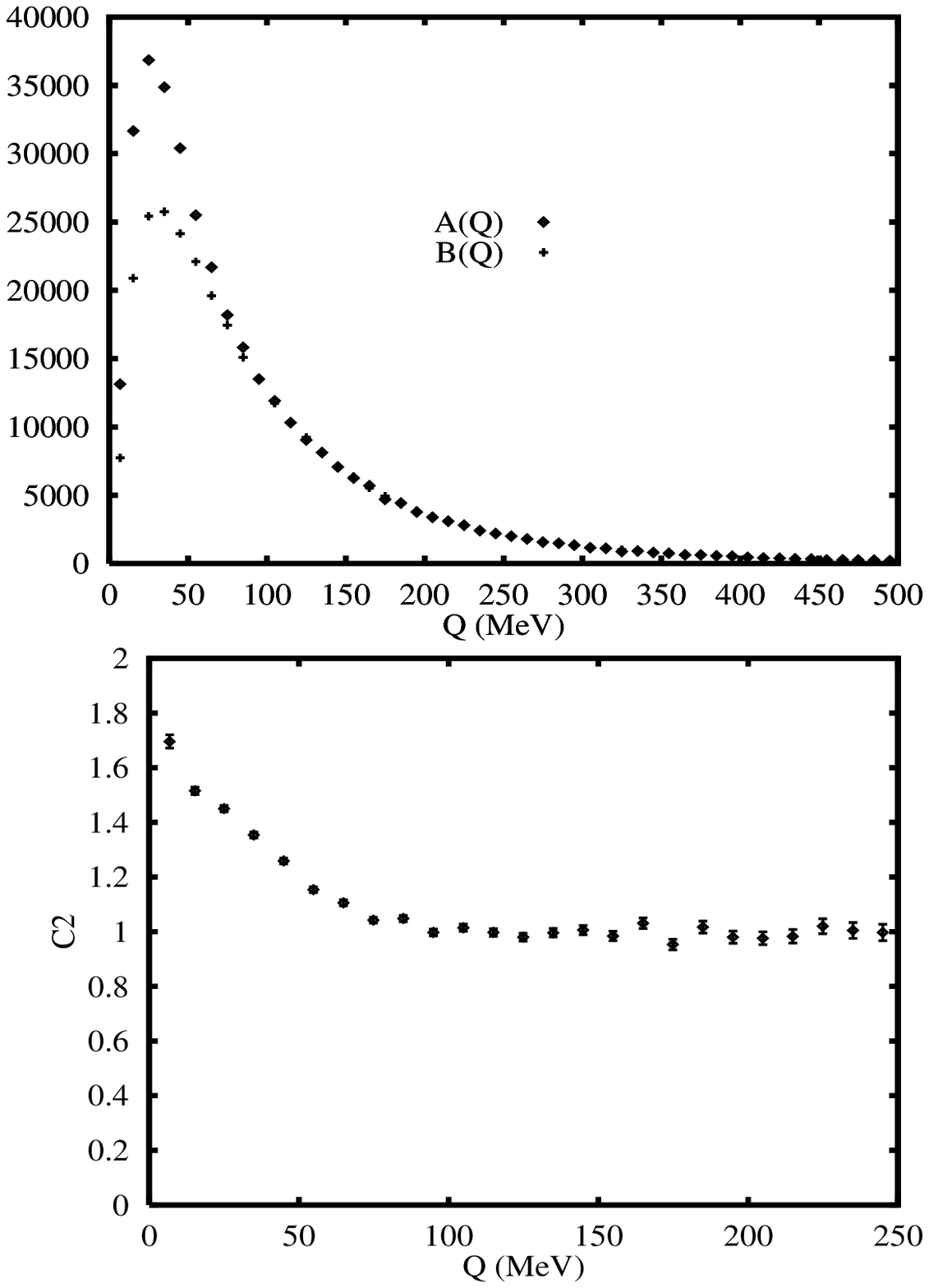}}
\end{center}
\vskip -5cm
\caption{
}
\label{mc1}
\end{figure}

\begin{figure}[t]
%
%
\begin{center}
\vskip -3cm
\hskip -2cm
          \leavevmode\epsfysize=9.0in
          {\epsfbox{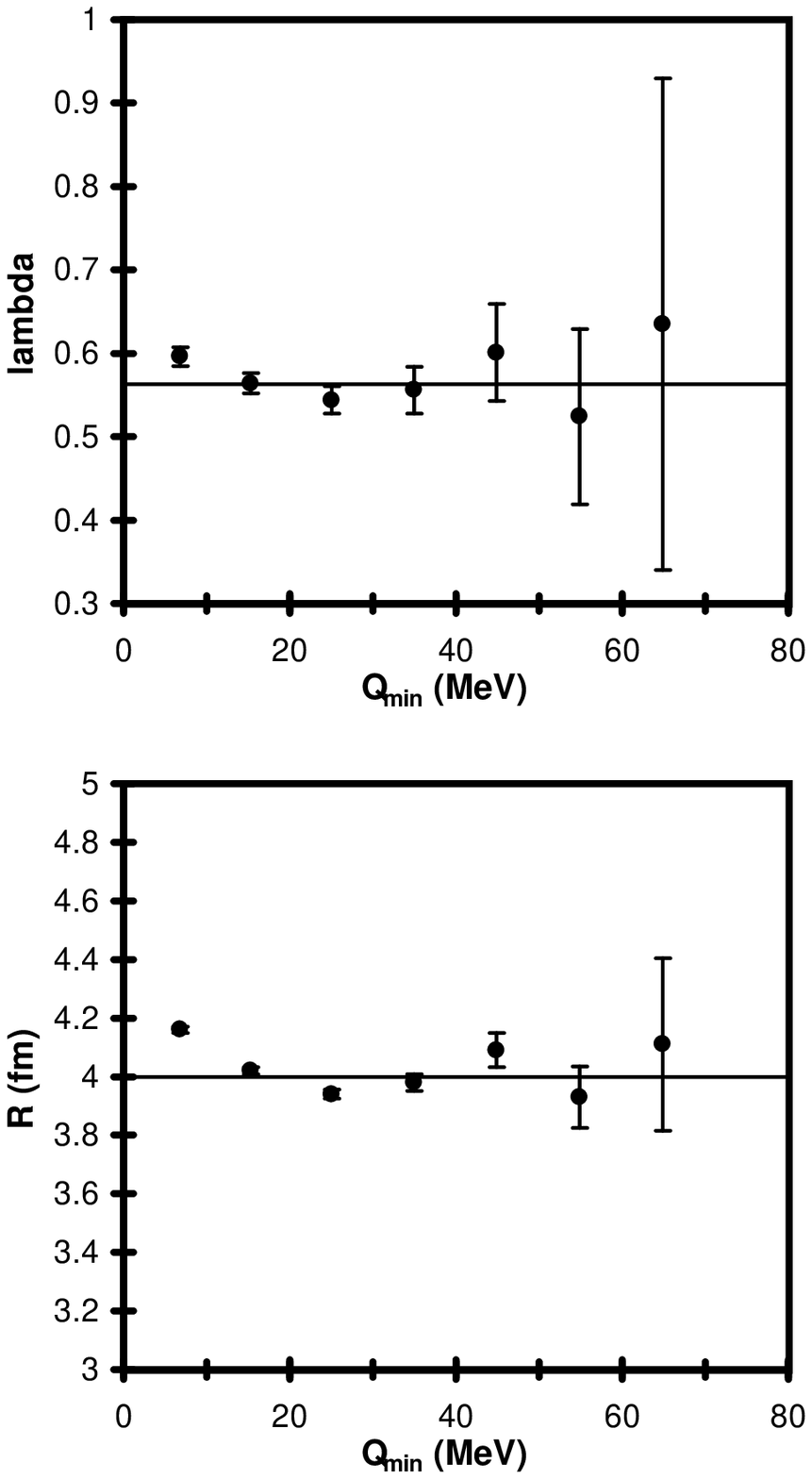}}
\end{center}
\vskip -5cm
\caption{
}
\label{mc2}
\end{figure}

\begin{figure}[t]
%
%
\centering
          \leavevmode\epsfysize=7.in
          {\epsfbox{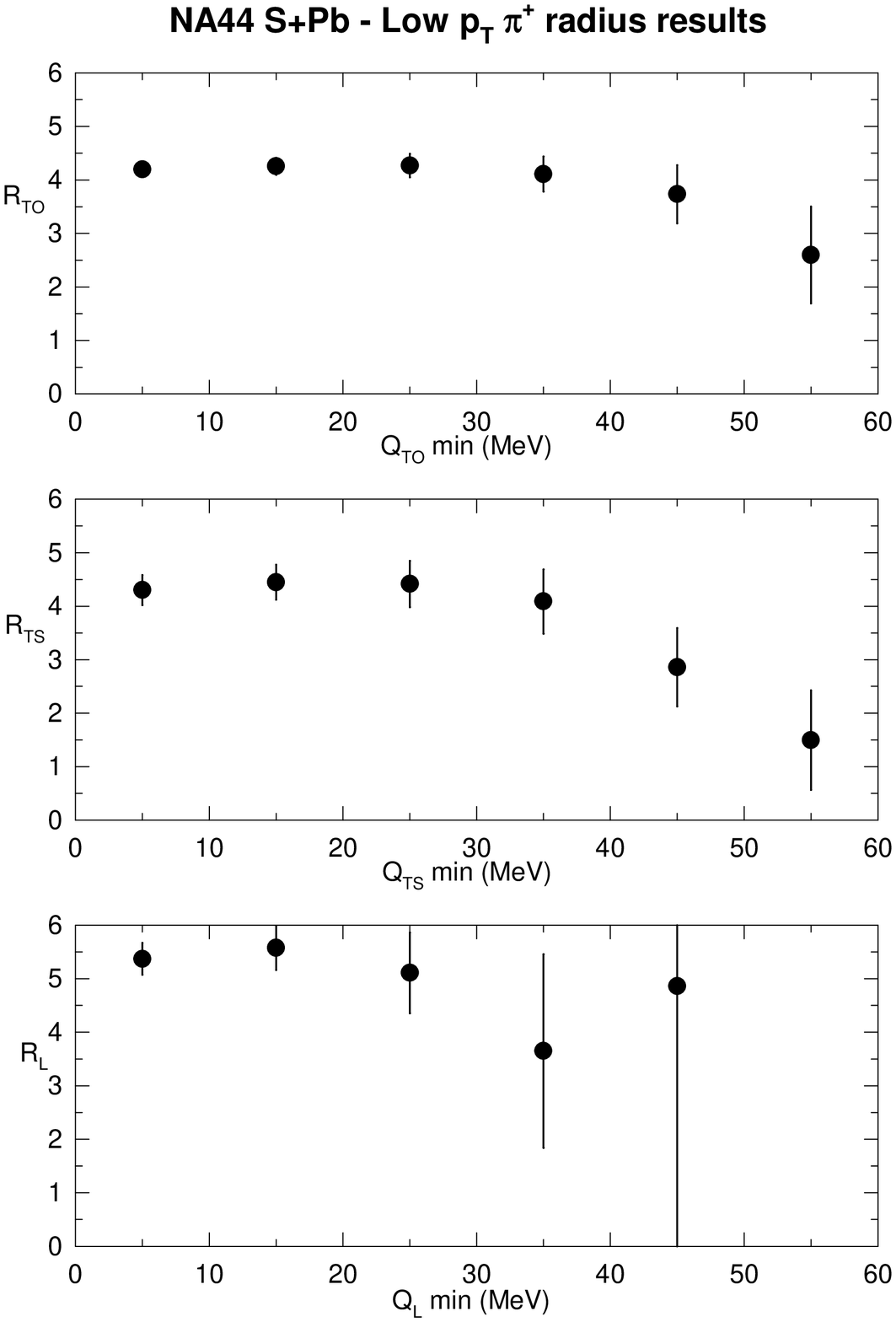}}
\caption{
	}
\label{f:lpr}
\end{figure}

\begin{figure}[t]
%
%
\centering
          \leavevmode\epsfysize=7.in
          {\epsfbox{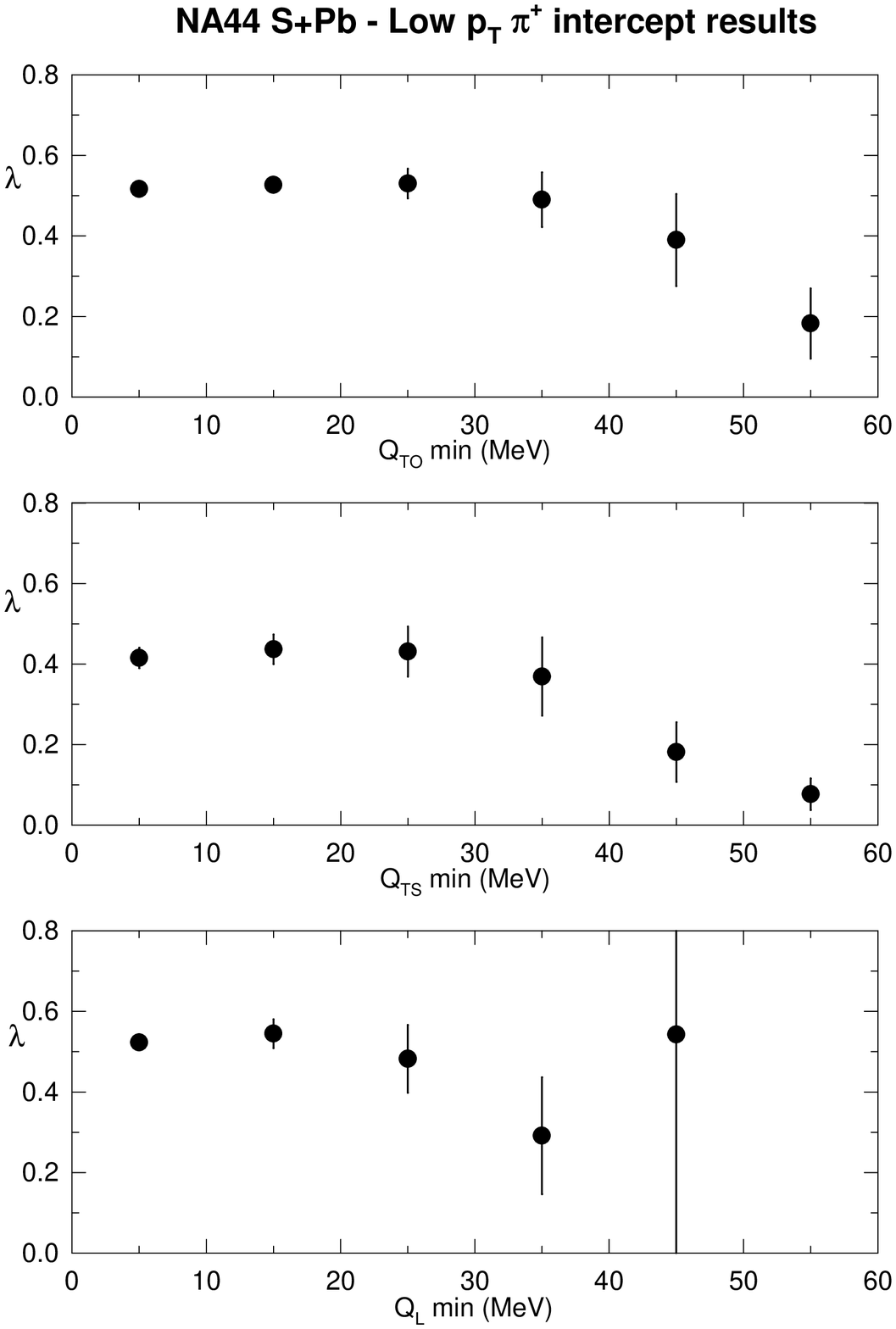}}
\caption{
	}
\label{f:lpi}
\end{figure}

\begin{figure}[t]
%
%
\centering
          \leavevmode\epsfysize=7.in
          {\epsfbox{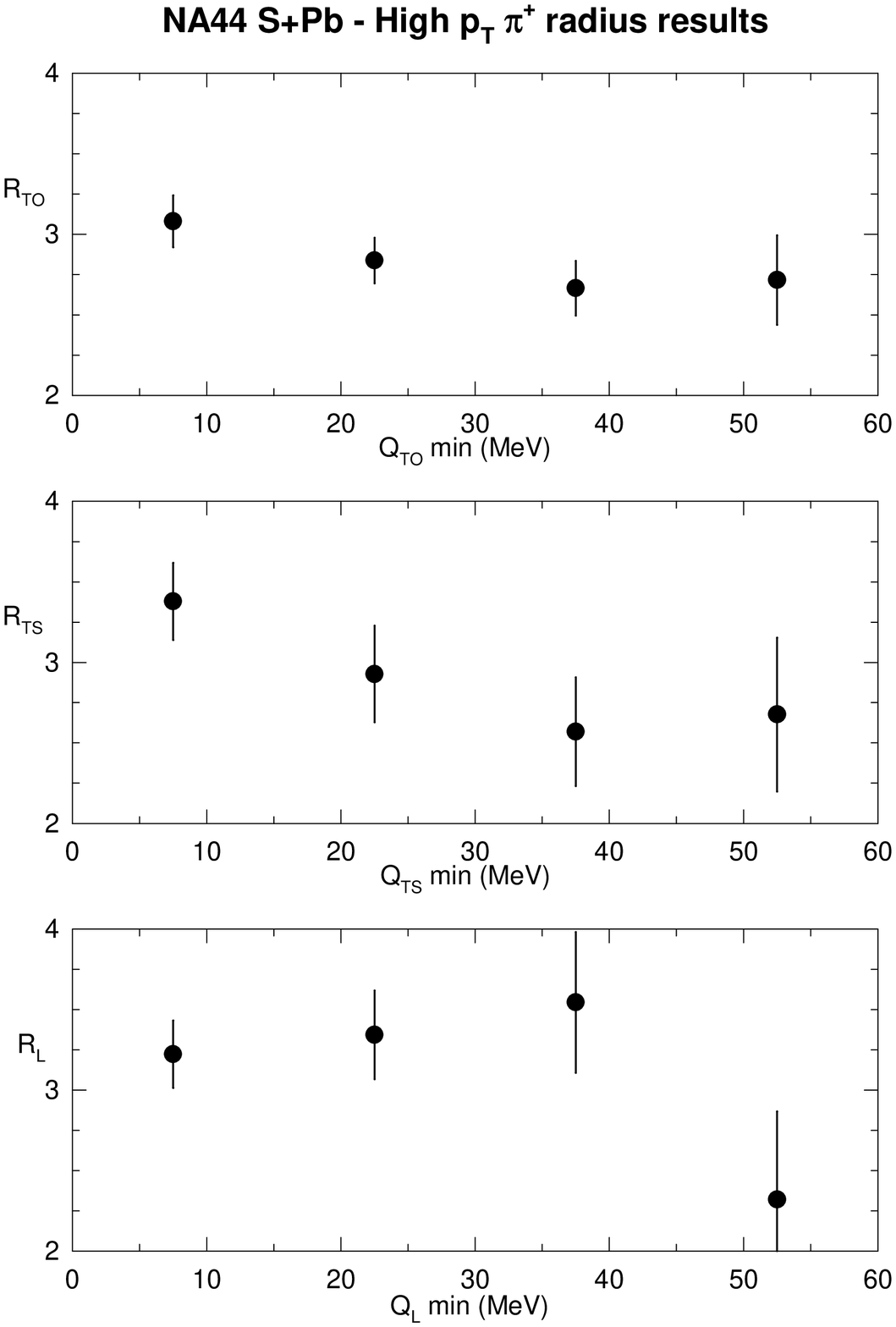}}
\caption{
	}
\label{f:hpr}
\end{figure}

\begin{figure}[t]
%
%
\centering
          \leavevmode\epsfysize=7.in
          {\epsfbox{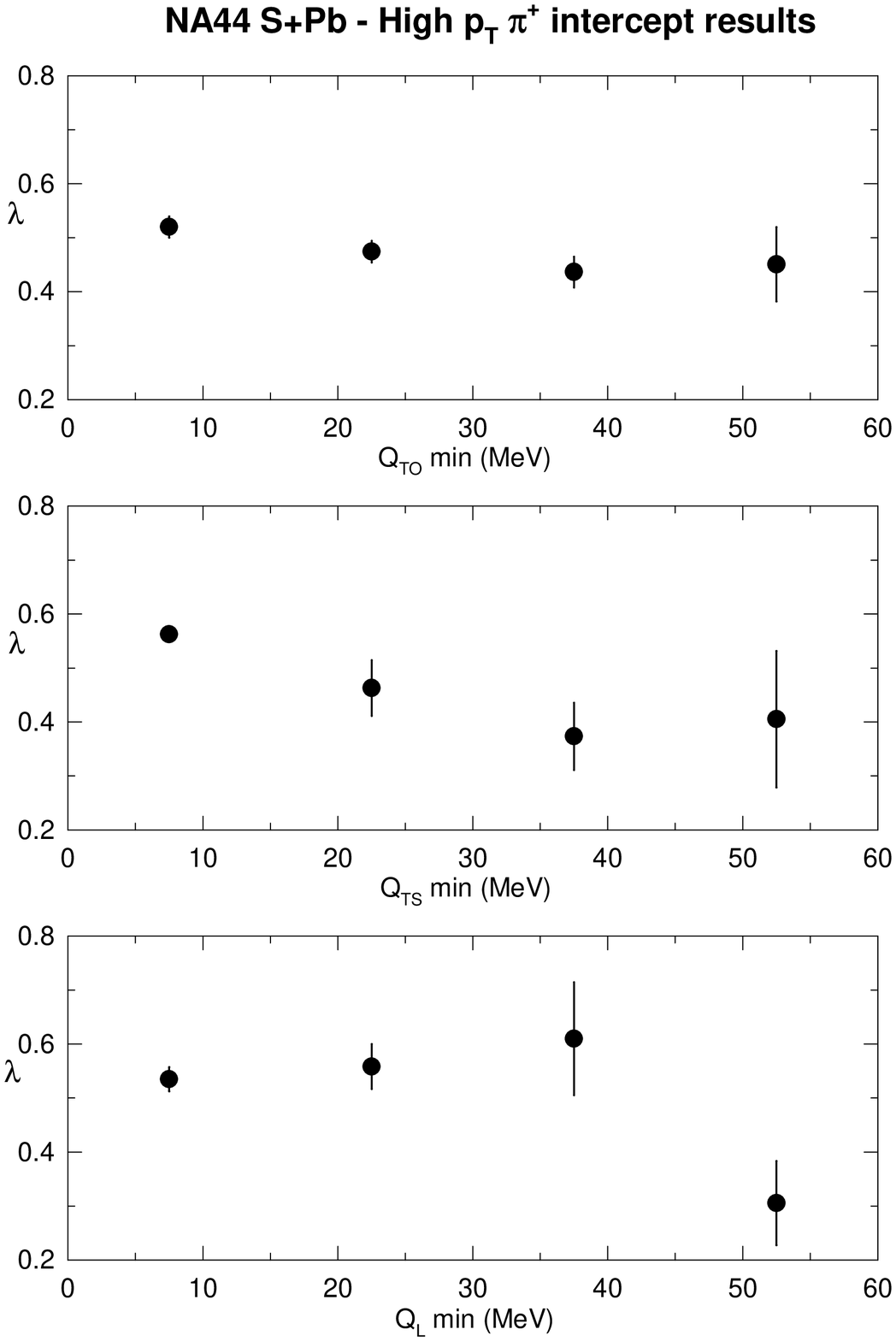}}
\caption{
	}
\label{f:hpi}
\end{figure}

\begin{figure}[t]
%
%
\centering
          \leavevmode\epsfysize=7.in
          {\epsfbox{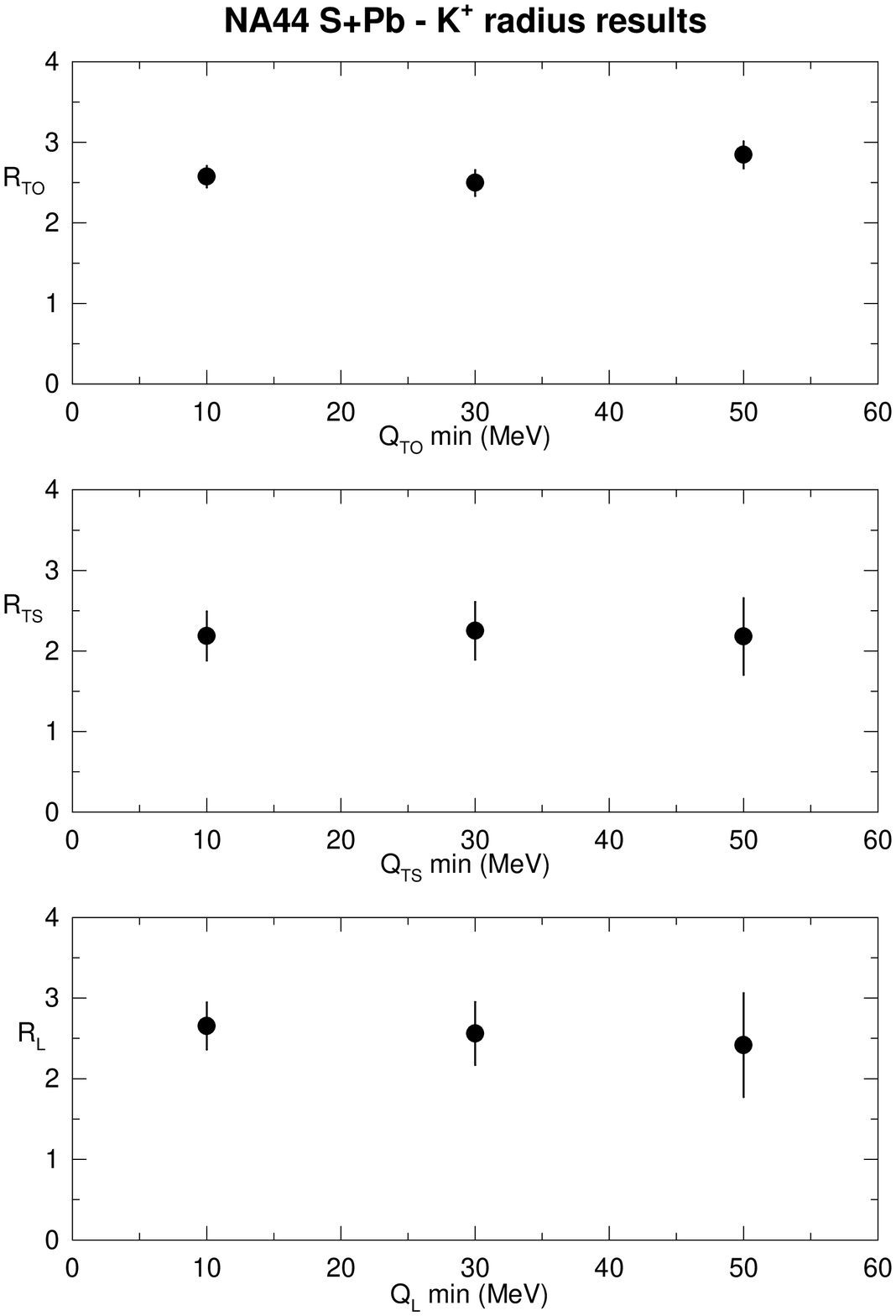}}
\caption{
	}
\label{f:kr}
\end{figure}

\begin{figure}[t]
%
%
\centering
          \leavevmode\epsfysize=7.in
          {\epsfbox{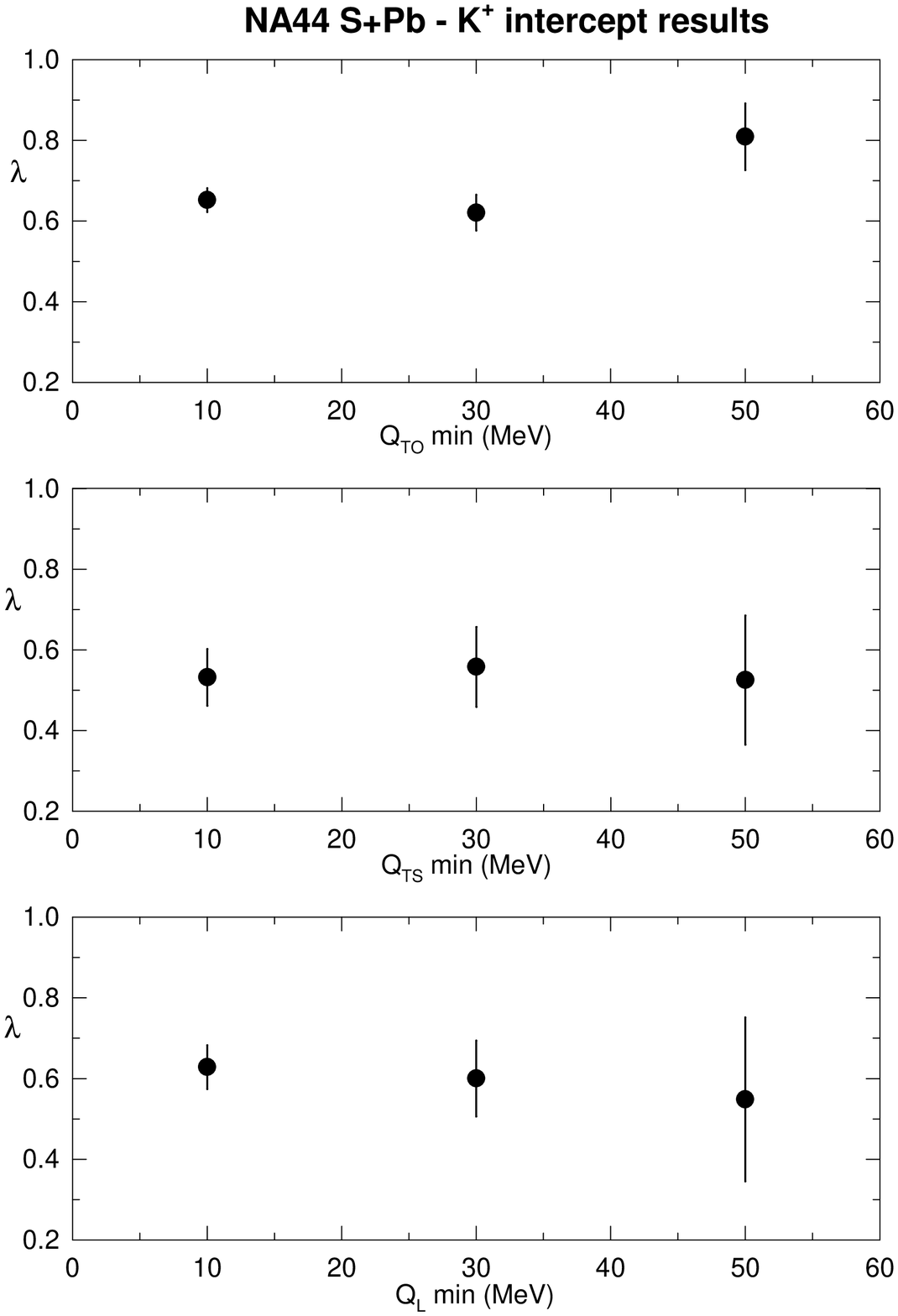}}
\caption{
	}
\label{f:ki}
\end{figure}

\begin{figure}[t]
%
%
\centering
          \leavevmode\epsfysize=3.5in
          {\epsfbox{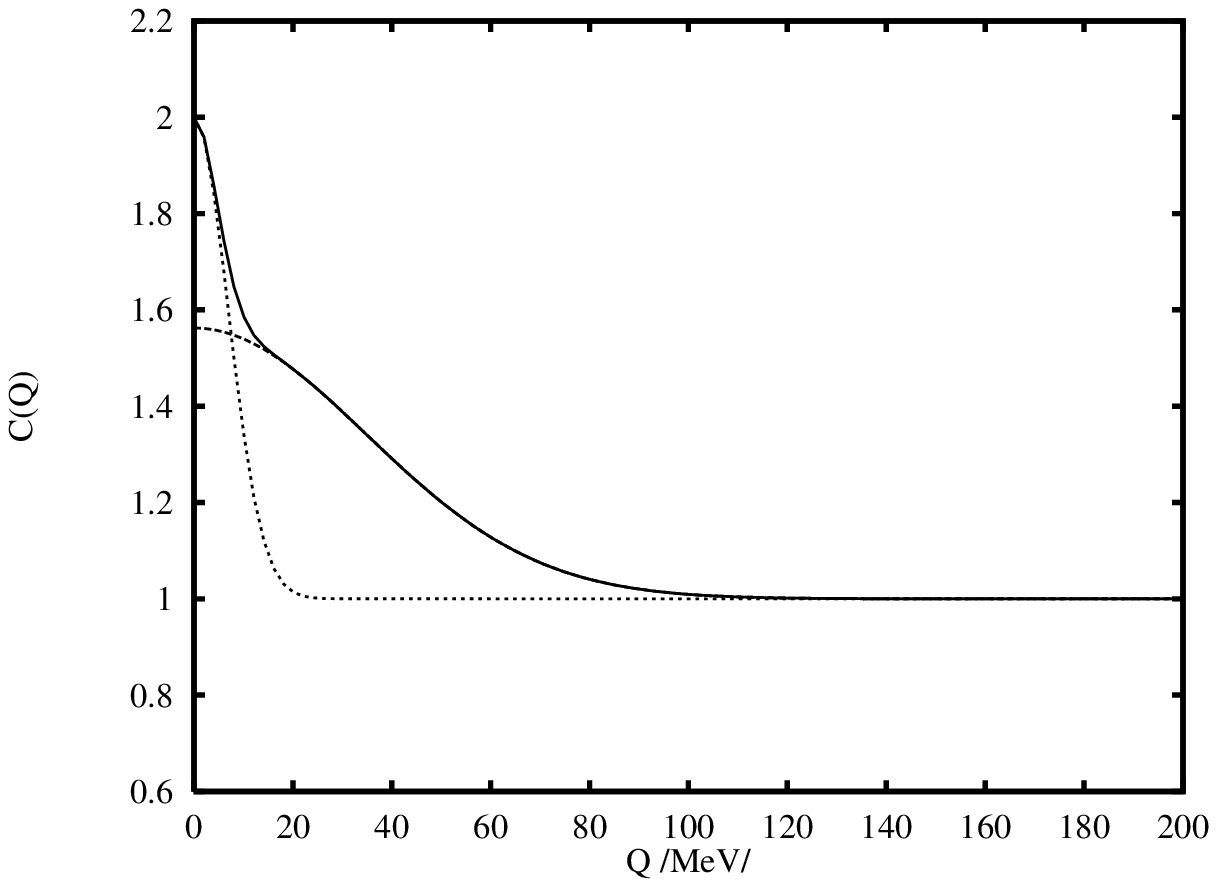}}
\caption{
	}
\label{f:append}
\end{figure}

	From the above example it is clear that a direct application of the 
of the original definitions of the Gaussian model-independent radii,
as was proposed originally in refs.~\cite{uli_s,uli_l} 
leads to {\it qualitatively 
and quantitatively unreliable} results if applied even to very simple models 
of core-halo type.
The Gaussian model-independent radii were obtained in
refs.~\cite{uli_s,uli_l} from an expansion around the $Q=0$  point.
However, according to Figure ~\ref{f:shaded}, in this region the correlation function
may have a narrow and unresolved structure, dominated by the
large regions of homogeneity in the halo.
Obviously, the variances as determined by the above relationships
shall be dominated by the variances of the halo part of the system,
	and in effect one obtains a nice Gaussian approximation to the 
Bose-Einstein correlation function --- {\it in the unresolved range}
of $Q < Q_{min} \approx 10$ MeV.
However, the resolved part of the correlation function shall be missed 
completely by this approximation.
See Figure~\ref{f:append} for an illustration of the effect.

From the above is also quite obvious, 
how to modify the original definitions
of the model-independent radii to get an approximation to the 
measured correlation functions:
the variances should be evaluated for the core emission function only.
	Although the possibility of such a characterization was discussed
	in ref.~\cite{uli-res,uli-summ} in a manner similar to the 
	present Appendix, the equations corresponding to 
	such a description were not given there.  
	We provide these below, since they may become a practically
	 useful tool to characterize the two-particle Bose-Einstein
	correlation functions in high energy heavy ion collisions.

This step results e.g. in a hydrodynamical like predictions for these
radii and in a $ 0 < \lambda_* < 1 $ intercept parameter of the correlation
function. A necessary condition for the applicability of the Gaussian
approximation is that the first two moments of the source distributions
of the core should be finite. The exceptional stable distributions (e.g.
the Lorentzian) do not satisfy this criteria. We know that the mean
of the core distribution is always finite hence the criteria is 
reduced to the requirement that the variances of the core be finite.
Under these conditions, the two-particle correlation function 
in the core/halo picture can be rewritten as
\bea
	C^{c/h}(\bk,\bdk ) & = & 1 + \lambda_*({\bf K}) \,
	\exp\left(- R^2_{i,j}({\bf K}) {\bf \Delta k}_i {\bf \Delta k}_j \right), \\
	\lambda_*({\bf K}) & = & [ N_{\bf c}({\bf K}) / N({\bf K}) ]^2, \\
	R_{i,j}^2({\bf K}) & = &
		 \langle (x_i - \bbet_i t) (x_j - \bbet_j t) \rangle_{\bf c}
		 - \langle (x_i - \bbet_i t)\rangle_{\bf c}
		   \langle (x_j - \bbet_j t) \rangle_{\bf c}, \\
	\langle f(x,{\bf p}) \rangle_{\bf c} & = & 
		\int d^4 x f(x,{\bf p})  S_{\bf c}(x,{\bf p}),
\eea
	where $i = side, out$ or $long$ as before, 
	and $S_{\bf c}(x,{\bf p})$ is the emission function 
	that characterizes the central core.
	Thus the halo contributes in this model to the reduction of the 
	intercept parameter only and the variances of the core
	correspond to the Gaussian core/halo model radii of the 
	measured correlation function. This result cannot be obtained
	with an expansion around $Q = 0$, on the other hand, it
	can be obtained with the help of a moment expansion of the 
	source distribution function around $x = \langle x \rangle_c$,
	which becomes possible if the second moment of the core distribution
	is finite.
	Hence, this  result corresponds to a large $Q$ expansion
	of the Bose-Einstein correlation function~\cite{darius}.

Although the above parameterization is a rather straight-forward
combination of core - halo model and the expressions for the
``model-independent" HBT radii, the results can still be considered 
simultaneously as a particular Gaussian approximation 
of the more general core/halo model result of eq.~(\ref{e:lamq})
as well as a generalization of the ``model-independent"
Gaussion HBT radii for systems of core/halo type.

\vfill\eject

\vfill\eject
\begin{center}
{\large\bf Figure Captions}
\end{center}
\begin{description}
\item[{\bf Fig. 1}]
	Full line indicates the effective, measured correlation 
	function of the core/halo model using Gaussian ansatz for both the 
	halo and the core, $R_c = 4.0$ fm/c, $R_H = 40$ fm/c.
	Dashed line stands for the full correlation function,
	which includes the effect from the halo, resulting in a narrow
	and unresolvable peak if $Q_{min} = 10$ MeV is the experimental resolution.
 	The extrapolated intercept parameter, $\lambda_*$ thus deviates from the 
	exact intercept of $\lambda_x = 1$, 
	however, $\lambda_*$ carries important information
	about the fraction of core particles at a mean momentum.
\item[{\bf Fig. 2}]
	The value of the three terms in the core-halo model correlation function.
	$f_c = 0.75, R_c = 4.0$ fm, $R_h = 40.$ fm
\item[{\bf Fig. 3}]
	The effect of $f_c$ on $C_2(Q)$ in the core-halo model for
	$R_c = 4.0$ fm and $R_h = 40.$ fm.
\item[{\bf Fig. 4}]
	The simulated $A(Q)$ and $B(Q)$ distributions, and the resulting
	$C_2(Q)$. Error bars are too small to be  seen for the actual and the
	background distributions $A(Q)$ and $B(Q)$. Input values
	were $R_c = 4.0$ fm, $R_h = 40.$ fm and  $f_c = 0.75$. 
	To simulate the $A(Q)$ distribution of the pion pairs,
	349,476 points were sampled, for the  $B(Q)$ distribution, 
	300,000. Note, that these $A(Q)$ and $B(Q)$ distributions were
	chosen to peak at 25 MeV, similarly to the CERN experiment NA44.
\item[{\bf Fig. 5}]
	The behaviour of the fitted $\lambda_*$ and $R$ parameters as  low $Q$
	data points are removed from $C_2(Q)$. 
	Solid line stands for $\lambda_* = f_c^2$ and $R_{fit} = R_{core}$.
\item[{\bf Fig. 6}]
	Radius parameters as a function of $Q_{min}$ for NA44's low $p_t$
	$\pi^+$ data.
\item[{\bf Fig. 7}]
	$\lambda_*$ parameter as a function of $Q_{min}$ for NA44's low $p_t$
	$\pi^+$ data.
\item[{\bf Fig. 8}]
	Radius parameters as a function of $Q_{min}$ for NA44's high $p_t$
	$\pi^+$ data.	
\item[{\bf Fig. 9}]
	$\lambda_*$ parameter as a function of $Q_{min}$ for NA44's high $p_t$
	$\pi^+$ data.
\item[{\bf Fig. 10}]
	Radius parameters as a function of $Q_{min}$ for NA44's $K^+$ data.
\item[{\bf Fig. 11}]
	$\lambda_*$ parameter as a function of $Q_{min}$ for NA44's $K^+$ data.
\item[{\bf Fig. 12}]
	Comparision of the full correlation function (full line)
 	to the core/halo model approximation (dashed line)
	and to the model-independent Gaussian approximation (dotted line).
	The model-independent radii yield qualitatively and quantitatively
	unreliable results if $\lambda_* < 1$.
\end{description}

\vfill\eject
\vfill
\vfill\eject

\vfill
\vfill\eject

\vfill
\vfill\eject

\vfill
\vfill\eject

\vfill
\vfill\eject

\vfill
\vfill\eject

\vfill

\vfill\eject

\vfill

\vfill\eject

\vfill

\vfill\eject

\vfill

\vfill\eject

\vfill

\vfill\eject

\vfill
\vfill\eject


\begin{thebibliography}{99}
\bibitem{HBT} 	R. Hanbury-Brown and R.Q. Twiss, 
		\emph{Nature} 178, 1046 (1956).
\bibitem{GGLP0} G. Goldhaber, W. B. Fowler, S. Goldhaber, 
		T. F. Hoang, T. E. Kalogeropolous and W. M. Powell, 
		Phys. Rev. Lett {\bf 3} (1959) 181.
\bibitem{GGLP} G. Goldhaber, S. Goldhaber, W. Lee, and A. Pais, 
		Phys. Rev.  \textbf{120}, 300 (1960).
\bibitem{gyu_ka}
	 M. Gyulassy, S. K. Kaufmann and L. W. Wilson,
         Phys. Rev. {\bf C20}, (1979) 2267
\bibitem{highpt} 
		H. Beker et al, Phys. Rev. Lett. \textbf{74}, 3340 (1995).
\bibitem{chalo} T. Cs\"{o}rg\H{o}, B. L\"{o}rstad, and J. Zim\'{a}nyi,
		hep-ph/9411307, Z. Phys. C {\bf 71} (1996) 491
\bibitem{nhalo} T. Cs\"org\H o, hep-ph/9705422, 
		Phys. Lett. B {\bf 409} (1997) 11
\bibitem{pratt_csorgo}
		S. Pratt, T. Cs\"org\H o and J. Zim\'anyi,
        	Phys. Rev. {\bf C42} (1990) 2646
\bibitem{zajc} 	William Zajc, 
		in \emph{Particle Production in Highly Excited Matter}, 
		ed. by H. Gutbord and J. Rafelski, NATO ASI series {\bf B303}
		(Plenum Press, New York, 1993) p. 435.
\bibitem{jzcst}	J. Zim\'anyi and T. Cs\"org\H o,
		hep-ph/9705432 (Phys. Rev. Lett. (1997) in preparation).
\bibitem{uli_l} S. Chapman, P. Scotto and U. Heinz, 
		Heavy Ion Physics {\bf 1} (1995) 1
\bibitem{rrr}
	I. Andreev, M. Pl\"umer, R. Weiner,
	Int. J. Mod. Phys. {\bf A8}(1993)4577; \\
	M. Biyajima et al, Progr. Theor. Phys. {\bf 84}(1990)931; \\
	N. Suzuki, M. Biyajima, Progr. Theor. Phys. {\bf 88} (1992) 609. 
\bibitem{bialas}
	A. Bialas, Acta Physica Polonica {\bf B23 } (1992) 561
\bibitem{bertsch}
	 G. F. Bertsch, Nucl. Phys. {\bf A498} (1989) 173c
\bibitem{lcms} 
		T. Cs\"org\H o and S. Pratt, Report No. KFKI-1991-28/A p. 75 
\bibitem{3d} 	T. Cs\"org\H o and B. L\"orstad, 
		Phys. Rev. C {\bf 54} (1996) 1390
\bibitem{RQMD}
	J. P. Sullivan et al, Phys. Rev. Lett. {\bf 70} (1993) 3000;\\
	R. D. Fields et al, Phys. Rev. {\bf C52} (1995) 986
\bibitem{res1} J. Bolz, U. Ornik, M. Pl\"{u}mer, B.R. Schlei, and R.M. Weiner,
		Phys. Lett. B {\bf 300}, 404 (1993).
\bibitem{res2}  J. Bolz, U. Ornik, M. Pl\"{u}mer, B.R. Schlei, and R.M. Weiner,
Phys. Rev. D {\bf 47}, 3860 (1993).
\bibitem{restable} 	
		H. Heiselberg, hep-ph/9602431, 
		Phys. Lett. B. {\bf 379} (1996) 27
\bibitem{res3}	B. R. Schlei et al, Phys. Lett. {\bf B376} (1996) 212
\bibitem{uli-summ} 
		U. Heinz, nucl-th/9609029,
		in {\it Correlations and Clustering
		Phenomena in Subatomic Physics}, Dronten, The Netherlands,
		Aug 4-18 1996 (NATO ASI Proc. Series, Plenum Press,
		to appear). 
\bibitem{uli-res}
		U. Heinz and U. A. Wiedemann,
		nucl-th/9611031
\bibitem{sinyu96}
		Yu. M. Sinyukov, S. V. Akkelin and Yu. A. Tolstkyh,
		Nucl. Phys. {\bf 610} (1996) 278c 
\bibitem{simon} 
		S. Nickerson, {\it ``A Halo Model of Heavy Ion Collisions",}
		M. Sc. Thesis, Dalhousie University, Halifax, Canada,
		June 1996
\bibitem{csorgo-kiang}
		T. Cs\"org\H o, S. Nickerson and D. Kiang,
		hep-ph/9611275
\bibitem{gyp96} S. S. Padula and M. Gyulassy,
		 Phys. Lett. B {\bf 348} (1995) 303 
\bibitem{afs} 	T. Akesson et al, AFS Collaboration,
		Phys. Lett. B {\bf 129} (1983) 269 ;\\
		Phys. Lett. B {\bf 187} (1987) 420 ;
		Z. Phys. C {\bf 36} (1987) 517.
\bibitem{na44hbt1} 
		H. Beker et al, NA44 Collaboration,
		Z. Phys. C {\bf 64} (1994) 209
\bibitem{na44hbt2} H. B{\o}ggild et al, NA44 Collaboration,
		Phys. Lett. B {\bf 302} (1993) 510.
\bibitem{lowpt} H. B{\o}ggild et al, Phys. Lett. B \textbf{349}, 386 (1995).
\bibitem{fax} 	B. L\"{o}rstad (private communication).
\bibitem{numrec} 	
		W.H. Press, B.P. Flannery, S.A. Teukolsky, and W.T. Vetterling,
		\emph{Numerical Recipes : The Art of Scientific Computing}, 
		(Cambridge University Press, New York, 1987).
\bibitem{na44npa} 
		M. Sarabura, Nucl. Phys. A \textbf{544}, 125c (1992).
\bibitem{dodd}  J. Dodd, NA44 Collaboration,
		Proc. XXV-th ISMPD, Stara Lesna, Slovakia, Sept. 1995.
		(D. Bruncko et al., eds, World Scientific, Singapore, 1996) 
\bibitem{uli_s} S. Chapman, P. Scotto, and U. Heinz, 
		Phys. Rev. Lett.  \textbf{74}, 4400 (1995).
\bibitem{kaon}	H. Beker et al, Z. Phys. C \textbf{64}, 209 (1994).
\bibitem{cross2} 
		T. Alber for the NA35/NA49 Collaborations, 
		Nucl. Phys. A \textbf{590} (1995) 453c. 
\bibitem{tect}  T. Cs\"org\H o, 
		talk at the HBT'96 conference,
		ECT*, Trento, Italy, September 1996.
\bibitem{darius}	D. Miskowiecz,
		talk at the HBT'96 conference,
		ECT*, Trento, Italy, September 1996.
\bibitem{vck}   S. E. Vance, T. Cs\"org\H o and D. Kharzeev, in preparation.
\end{thebibliography}
\end{document}